\documentclass{IEEEtran}
\usepackage{geometry}
\usepackage{graphics}
\usepackage{graphicx}
\graphicspath{{./fig/}}
\usepackage{color}
\usepackage{subfig}
\usepackage{amssymb}
\usepackage{amsmath}
\usepackage{mathtools}
\usepackage{amsfonts}
\usepackage{ntheorem}
\usepackage{enumitem} 
\usepackage{mathrsfs}
\usepackage{latexsym, bm}
\usepackage{hyperref}
\hypersetup{
  colorlinks   = true, 
  urlcolor     = blue, 
  linkcolor    = blue, 
  citecolor   = red 
}

\usepackage{pdfpages}
\newtheorem{theorem}{Theorem}

\newtheorem{definition}{Definition}
\newtheorem{lemma}{Lemma}
\newtheorem{corollary}{Corollary}
\newtheorem{conjecture}{Conjecture}
\newtheorem{remark}{Remark}

\newtheorem{assumption}{Assumption}

\newtheorem*{keywords}{Keywords}

\newcommand{\thetamax}{\theta_{\max}}
\newcommand{\thetamin}{\theta_{\min}}
\newcommand{\thetaave}{\theta_{\textup{ave}}}
\newcommand{\Vfully}{\mathcal{V}_{\textup{f}}}
\newcommand{\Vpartly}{\mathcal{V}_{\textup{p}}}

\newcommand{\thetai}[1]{\theta_{#1}}
\newcommand{\interior}{\operatorname{int}}
\newcommand{\norm}[1]{\| #1 \|}

\DeclareMathOperator{\diag}{diag}
\usepackage{lineno}

\newcommand{\until}[1]{\{1,\dots, #1\}}

\newcommand{\setdef}[2]{\{#1 \; | \; #2\}}

\newcommand{\map}[3]{#1: #2 \rightarrow #3}

\newcommand\oprocendsymbol{\hbox{$\square$}}
\newcommand\oprocend{\relax\ifmmode\else\unskip\hfill\fi\oprocendsymbol}

\DeclareSymbolFont{bbold}{U}{bbold}{m}{n}
\DeclareSymbolFontAlphabet{\mathbbold}{bbold}

\usepackage[prependcaption,colorinlistoftodos]{todonotes}

\begin{document}




\title{Social power evolution in influence networks with stubborn
  individuals}

\author{Ye Tian, Peng Jia, Anahita Mirtabatabaei, Long Wang, Noah
  E. Friedkin, Francesco Bullo, \IEEEmembership{Fellow, IEEE} \thanks{The
    research of Y. Tian and L. Wang was supported by National Natural Science
    Foundation of China (61751301 and 61533001). Y. Tian was also supported
    by the China Scholarship Council (CSC No. 201706960062). The research
    of N. E. Friedkin and F. Bullo was supported in part by the U. S. Army
    Research Laboratory and the U. S. Army Research Office under grant
    number W911NF-15-1-0577.  {\it(Corresponding author: Long Wang)}}

\thanks{Y. Tian is with the Center for Complex Systems, School of Mechano-electronic Engineering, Xidian University, Xi'an 710071, China, the Mechanical Engineering Department and the Center of Control, Dynamical Systems and Computation, UC Santa Barbara, CA 93106, USA.
{\tt\small tinybeta7.1@gmail.com}}
\thanks{P. Jia, A. Mirtabatabaei and F. Bullo are with the Mechanical Engineering Department and the Center of Control, Dynamical Systems and Computation, UC Santa Barbara, CA 93106, USA.
 {\tt\small\{pengjwt,anaheata\}@gmail.com; bullo@engineering.ucsb.edu}}
\thanks{L. Wang is with the Center for Systems and Control, College of Engineering, Peking University, Beijing 100871, China. 
{\tt\small longwang@pku.edu.cn}}
\thanks{N. E. Friedkin is with the Department of Sociology and the Center for Control, Dynamical Systems, and Computation, UC Santa Barbara, CA 93106, USA. 
{\tt\small friedkin@soc.ucsb.edu}}
}

\maketitle

\begin{abstract}
This paper studies the evolution of social power in influence networks with stubborn individuals. Based on the Friedkin-Johnsen opinion dynamics and the reflected appraisal mechanism, two models are proposed over issue sequences and over a single issue, respectively. These models generalize the original DeGroot-Friedkin (DF) model by including stubbornness. To the best of our knowledge, this paper is the first attempt to investigate the social power evolution of stubborn individuals basing on the reflected appraisal mechanism. Properties of equilibria and convergence are provided. We show that the models have same equilibrium social power and convergence property, where the equilibrium social power depends only upon interpersonal influence and individuals' stubbornness. Roughly speaking, more stubborn individual has more equilibrium social power. Moreover, unlike the DF model without stubbornness, we prove that for the models with stubbornness, autocracy can never be achieved, while democracy can be achieved under any network topology.
\end{abstract}

\begin{keywords}
Opinion dynamics, influence networks, social power, reflected appraisal, dynamical systems, mathematical sociology
\end{keywords}

\section{Introduction}\label{s1}
\paragraph*{Problem description and motivation}
This paper investigates the evolution of social power in influence networks with stubborn individuals. Two models are formulated over issue sequences and over a single issue, respectively. The first model incorporates the Friedkin-Johnsen (FJ) opinion dynamics and the reflected appraisal mechanism to characterize the process of opinion change on each issue and evolution of social power over issue sequences, respectively. The second model is a variation of the first model, in which the processes of opinion dynamics and reflected appraisal take place on a single issue. In the DeGroot-Friedkin (DF) model, the process of opinion dynamics is described by the DeGroot model, where individuals are completely open to interpersonal influence. However, it has been shown by empirical evidence that the FJ model is more realistic and predictive in modelling opinion changes. This paper extends the original DF model by including stubbornness. Rigorous analysis and numerical experiments are provided for equilibria and convergence properties. We aim to uncover the difference between the evolution of social power in groups with and without stubbornness. 

\paragraph*{Literature review}
The investigation of social networks has attracted much attention from  applied mathematics, sociology, control theory and economics, etc., over the last several decades. Classic dynamic models of interest concern on how individuals exchange and integrate opinions on a certain issue \cite{AVP-RT:17}, \cite{AVP-RT:18}, including the DeGroot model \cite{JRPF:56}, \cite{FH:59}, \cite{MHDG:74}, the Abelson model \cite{RPA:64}, the FJ model \cite{NEF-ECJ:99} and the Hegselmann-Krause model \cite{RH-UK:02}, \cite{AM-FB:11f}, to name but a few. In this literature, the FJ model, which generalizes the DeGroot model by introducing stubbornness, is particularly of interest, due to its predictive ability in human-subject experiments \cite{NEF-FB:16d}, \cite{NEF-PJ-FB:14n}, \cite{NEF-ECJ:11}, \cite{NEF-ECJ:99}. Further investigations of the FJ model include \cite{NEF-AVP-RT-SEP:16}, \cite{YT-LW:18}, \cite{SEP-AVP-RT-NEF:17} and the references therein.

Recently, the evolution of social power, namely, the amount of influence or relative control of individuals during opinion discussion, has drawn considerable interest. The study of social power dynamics was initiated by Friedkin \cite{NEF:11} with a mathematization of the psychological mechanism of reflected appraisal. A rigorous mathematical model and dynamical system analysis was provided by Jia et al.~\cite{PJ-AM-NEF-FB:13d}, known as the DF model, which integrates, respectively, the DeGroot model and the reflected appraisal to describe the opinion dynamics on each issue and the social power evolution over issue sequences. Empirical evidence in support of the reflected appraisal mechanism was provided in \cite{NEF-PJ-FB:14n}. 

Several extensions and variations of the DF model has been presented since its introduction. Jia et al. \cite{PJ-NEF-FB:14m} extended it to the case that the relative interaction matrix is reducible. A single timesacle DF model was proposed and investigated in \cite{PJ-NEF-FB:14e}, where reflected appraisal  and opinion dynamics take place on a single issue. A modified DF model was proposed in \cite{ZX-JL-TB:15}, where social power is unpdated before opinion consensus. A novel stability analysis method for nonlinear Markov chains formulated on the DF model was provided in \cite{ZA-RF-AH-YC-TTG:17}. Chen et al. \cite{GC-XD-NEF-FB:16j} extended the DF model to the scenario where the relative interaction matrix is switching and stochastic. In \cite{MY-JL-BDOA-CY-TB:17b}, Ye et al. extended the DF model to the setting that the relative interaction matrix is switching in a finite set, an approach based on nonlinear contraction analysis \cite{WL-JJES:98} was employed to address the convergence properties. 

\paragraph*{Contributions}
This paper extends the DF model by including stubbornness. First, we
propose two models on social power evolution of stubborn individuals over
issue sequences and over a single issue, respectively. These models cover
two prevalent scenarios in the practice. That is, for specially designed
groups, it is feasible or necessary to appraise each member's performance
or importance after discussion on each issue; for loosely assembled or
spontaneously arisen groups, discussion on issues may be persistent and
reflected appraisal may take place after each opinion update.

Second, we study the properties of equilibria. We prove that for the two models, equilibrium social power is equivalent. Based on the equivalence, we derive the condition for uniqueness of equilibrium social power under general topology, and provide lower and upper bounds for the equilibrium social power. Moreover, we analyze the relationship between equilibrium social power and stubbornness, interpersonal influence, respectively. A sufficient and necessary condition for the existence of democratic equilibrium social power is also provided. For the case that the influence network is star topology, we analyze the uniqueness of the equilibrium social power in the settings that the center node is fully stubborn and partially stubborn, respectively. In the former case, we prove that the center node occupies the largest equilibrium social power, while the ordering of the equilibrium social power of partially stubborn individuals is consistent with the ordering of their stubbornness. In the later case, we show that individuals' social power at equilibrium increases as their stubbornness or influence weights accorded by center node increase. 

Third, we establish the convergence properties. For the model over issue sequences, we prove that all its trajectories globally converge to the unique equilibrium exponentially fast. The convergence properties under star topologies with fully stubborn and partially stubborn center node are also provided, respectively. For the model over a single issue, we prove that individuals' social power globally exponentially converges to the unique equilibrium if their stubborn levels are higher than $1/2$. Moreover, in the case that the relative interaction matrix is doubly-stochastic and individuals are uniformly stubborn, we prove that individuals' social power globally exponentially converges to the democratic social power structure. Finally, based on the simulation results and the Chernoff bound, we provide a conjecture for the uniqueness and global attractivity of the equilibrium. 

Our investigation reveal some findings which are of sociological interest. First, the equivalence of equilibrium social power implies that the reflected appraisal mechanism is robust with respect to variations in the time scales at which opinions and social power evolve. Second, individuals will forget their initial social power exponentially fast, and the equilibrium social power only depends on interpersonal influence and stubbornness. Third, the social power of stubborn individual can never be $0$, which means that stubbornness leads to social power. Moreover, for individuals embedded in symmetric influence networks or accorded same influence weights by partially stubborn individuals, more stubbornness leads to more social power. Finally, in groups consisting of stubborn individuals, autocratic social power never emerges, while democratic social power can be achieved regardless of the network topology. From this perspective, stubbornness enables groups to prevent the emergence of autocracy and to achieve democracy.

Lastly, compared with our preliminary conference paper \cite{AM-PJ-NEF-FB:13x}, this article contains several results and updates not found in \cite{AM-PJ-NEF-FB:13x}. First, we propose a new model on the social power evolution of stubborn individuals over a single issue, and analyze properties of equilibria and convergence. Second, for the model over issue sequences, we derive a milder condition for the uniqueness of equilibrium social power, and provide convergence analysis, which is not addressed in \cite{AM-PJ-NEF-FB:13x} except the case that the influence network is doubly-stochastic and individuals are homogeneous. Third, we discuss the properties of the equilibrium social power and its relationship with the influence network and individuals' stubbornness.  

\paragraph*{Paper organization}
In Section \ref{s2}, we propose the DF model with stubbornness over issue sequences and over a single issue, respectively. In Section \ref{s3}, properties of the equilibrium social power is analyzed. In Section \ref{s4}, we establish the convergence properties. Simulations and a conjecture are provided in Section \ref{s5}. Section \ref{s6} concludes the paper and all proofs are in Appendices.
\paragraph*{Notations}
Let $\mathbf{1}_{n}$ and $I_{n}$ denote the $n\times 1$ all-ones vector and the $n\times n$ identity matrix, respectively. $\mathbf{e}_{i}$ denotes the $i$-th standard basis vector with proper dimension. Given $\delta\in\mathbb{R}^{n}$, $\diag(\delta)$ denotes a diagonal matrix with diagonal elements $\delta_{1}, \delta_{2}, \dots, \delta_{n}$. The $n$-simplex is denoted by $\Delta_{n}\!=\!\setdef{x\in\mathbb{R}^{n}}{x\geq 0, \mathbf{1}^{T}_{n}x\!=\!1}$. Its interior is denoted by $\interior{\Delta_{n}}\!=\!\setdef{x\in\mathbb{R}^{n}}{x> 0, \mathbf{1}^{T}_{n}x\!=\!1}$. A nonnegative matrix is row-stochastic (column-stochastic) if its row (column) sums are $1$; it is doubly-stochastic if both its row and column sums are $1$. The weighted digraph $\mathcal{G}(W)$ associated to nonnegative matrix $W$ is defined as: the node set is $\until n$; there is a directed edge $(i,j)$ from nodes $i$ to $j$ if and only if $W_{ij}>0$. $\mathcal{G}(W)$ is a star topology if all its directed edges are either from or to a center node. A strongly connected component (SCC) of $\mathcal{G}(W)$ is a maximal strongly connected subgraph. A SCC is called sink SCC if there exists no directed edge from this SCC to others. 

\section{Modeling}\label{s2}
In this section, we propose two models describing the social power evolution of stubborn individuals over issue sequences and over a single issue, respectively.
\subsection{The DF model with stubborn individuals over issue sequences}\label{s2.1}
Consider $n\!\geq\!2$ individuals discussing a sequence of issues $s=0, 1, 2, \dots$ in an influence network formulated by weighted digraph $\mathcal{G}(C)$, where $C$ is the row-stochastic, zero-diagonal relative interaction matrix. Let $y_{i}(s,k)\in\mathbb{R}$ denote the opinion of individual $i$ on issue $s$ at time $k$. $\theta_{i}\in[0,1]$ denotes individual $i$'s susceptibility to interpersonal influence, i.e., $1-\theta_{i}$ represents its stubbornness to initial opinion. Assume that during the discussion of issue $s$, the self-appraisal of individual $i$, denoted by $x_{i}(s)\in[0,1]$, is static, and each individual forms its opinion according to the FJ model
\begin{equation*}
y_{i}(s,k+1)=\theta_{i}\sum_{j=1}^{n}W_{ij}(s)y_{j}(s,k)+(1-\theta_{i})y_{i}(s,0).
\end{equation*}
Assume that $W_{ii}(s)=x_{i}(s)$, and $W_{ij}(s)=(1-x_{i}(s))C_{ij}$, i.e., individuals' self-weights are equal to their self-appraisals. Let $y(s,k)$ and $\theta$ denote the vectors of individuals' opinions and susceptibilities, we have 
\begin{equation}\label{e1}
y(s,k+1)=\Theta W(x(s)) y(s,k)+(I_{n}-\Theta)y(s,0),
\end{equation}
where $\Theta\!=\!\diag(\theta)$, $x(s)\in\Delta_{n}$, and $W(x(s))\!=\!\diag(x(s))+(I_{n}-\diag(x(s)))C$. 

\begin{assumption}\label{A1}
Suppose that every sink SCC of $\mathcal{G}(C)$ has at least one stubborn individual, and $\theta_{i}<1$ if $x(0)=\mathbf{e}_{i}$.
\end{assumption}

Assumption~\ref{A1} ensures that the FJ opinion dynamics converges on each issue. By Lemma III.1 in \cite{AM-PJ-NEF-FB:13x}, $\Theta W(x(s))$ is strictly row-substochastic for any $s\geq 0$ under Assumption~\ref{A1}. Hence, on each issue $s$, there holds
\begin{equation}\label{e2}
y(s,\infty)=V(x(s))y(s,0),
\end{equation}
where $V(x(s))=(I_{n}-\Theta W(x(s)))^{-1}(I_{n}-\Theta)$ is row-stochastic.

Equation (\ref{e2}) implies that each individual's opinion converges to a convex combination of all individuals' initial opinions. In other words, $V_{ij}(s)$ is the influence of individual $j$'s initial opinion to individual $i$'s final opinion on issue $s$. Consequently, $(1/n)\sum_{i=1}^{n}V_{ij}(s)$, which represents individual $j$'s relative control on other individuals' final opinions, is individual $j$'s social power exerted on issue $s$, as defined in \cite{DC:59}. According to the reflected appraisal mechanism \cite{NEF:11}, individuals' self-appraisals on each issue are set equal to their social power they exerted over prior issue. That is,
\begin{equation}\label{e3}
x(s+1)=V(x(s))^{T}\dfrac{\mathbf{1}_{n}}{n}.
\end{equation}
Since $V(x(s))$ is row-stochastic, equation \eqref{e3} ensures that $x(s+1)\in\Delta_{n}$. 
\begin{definition}\label{d1}
(The DeGroot-Friedkin model with stubborn individuals over issue sequences) Consider an influence network with $n\!\geq\!2$ individuals discussing a sequence of issues $s=0,1,2,\dots$ Suppose that Assumption~\ref{A1} holds. Let $C$ and $\Theta\!=\!\diag(\theta_{1},\theta_{2},\dots,\theta_{n})$ be the row-stochastic, zero-diagonal relative interaction matrix and the diagonal matrix representing individuals' susceptibilities, respectively. Then, the DeGroot-Friedkin model with stubborn individuals over issue sequences is
\begin{equation}\label{e4}
\begin{split}
x(s+1)=(I_{n}-\Theta)(I_{n}-W(x(s))^{T}\Theta)^{-1}\dfrac{\mathbf{1}_{n}}{n},
\end{split}
\end{equation}
where $W(x(s))\!=\!\diag(x(s))+(I_{n}-\diag(x(s)))C$.
\end{definition}
Define $\map{F}{\Delta_{n}}{\Delta_{n}}$ as
\begin{equation}\label{e7}
F(x)\!=\!(I_{n}-\Theta)(I_{n}-W(x)^{T}\Theta)^{-1}\dfrac{\mathbf{1}_{n}}{n}.
\end{equation}
Then, system \eqref{e4} can be written as $x(s+1)=F(x(s))$.

System (\ref{e4}) generalized the original DF model to the case that individuals are anchored to their initial opinions during the discussion of each issue. Empirical evidence supporting this generalization is provided in \cite{NEF-ECJ:99}, \cite{NEF-PJ-FB:14n} and \cite{NEF-FB:16d}, which substantiate that the presence of stubbornness is prevalent in human-subject experiments, and the model including stubbornness is more predictive. Note that if $\Theta=I_{n}$, then system (\ref{e4}) is the original DF model. Whereas, at the presence of stubbornness, individuals' final opinions on each issue depend not only on the relative influence network, but also on their stubbornness, and generally can not achieve consensus \cite{YT-LW:18}. This is different from the original DF model, in which individuals' social power can be captured by the dominant left eigenvector of $C$ under the assumption that all sink SCCs of $\mathcal{G}(C)$ are aperiodic. 

According to Definition \ref{d1}, for any $s>0$ and $x(0)\in\Delta_{n}$, if $\theta_{i}\!=\!1$, then $x_{i}(s)\!\equiv\!0$; if $\theta_{i}\!=\!0$ for all $i$, then $x(s)\!\equiv\!\mathbf{1}_{n}/{n}$. For simplicity, we have the following assumption. 
\begin{assumption}\label{A2}
Suppose that $\theta_{i}<1$ for any $i\in\until n$, and there exists at least one individual $j$ with $\theta_{j}>0$. 
\end{assumption}
Note that Assumption~\ref{A2} implies Assumption~\ref{A1}.

\begin{remark}\label{r1}
 In model~(\ref{e4}) individual's relative control over the prior
 discussion is appraised by computing $(I_{n}-W(x(s))^{T}\Theta)^{-1}$ and
 by averaging the columns of $V(x(s))$; both steps are unrealistic for an
 individual to perform in a large group because of information and
 computational requirements.  Here we propose a simple distributed
 dynamical process by which individuals can perceive their social power by
 using the local interpersonal influence information.  Assume that
 each individual knows the group size $n$, the susceptibilities of
 individuals who accord interpersonal influence to it and the accorded
 influence weights. At each issue $s$ and time $k$, let $p_{i}(s,k)$ denote
 the perceived social power of individual $i$, $W(s)=\diag(x(s))+(I-\diag(x(s)))C$ denote the influence matrix. Then, individual $i$ perceives its social power during the discussion of issue $s$ according to
\begin{equation*}
p_{i}(s,k+1)=(1-\thetai{i})\sum_{j=1}^{n}\dfrac{\thetai{j}W_{ji}(s)p_{j}(s,k)}{1-\thetai{j}}+\dfrac{1-\thetai{i}}{n}.
\end{equation*}
That is,
\begin{equation*}
\begin{split}
p(s,k+1)=\tilde{W}(s)p(s,k)+(I_{n}-\Theta)\dfrac{\mathbf{1}_{n}}{n},
\end{split}
\end{equation*}
where $\tilde{W}(s)=(I_{n}-\Theta)W(s)^{T}\Theta(I_{n}-\Theta)^{-1}$, whose spectral radius is strictly less than $1$ under Assumption~\ref{A2}. Hence, $p(s,\infty)=(I_{n}-\Theta)(I_{n}-W(s)^{T}\Theta)^{-1}\mathbf{1}_{n}/{n}=x(s+1)$ for any $p(s,0)\in\mathbb{R}^{n}$.
\end{remark}

\subsection{The DF model with stubborn individuals over a single issue}\label{s2.2}

We now propose a variation of model (\ref{e4}), in which the processes of
reflected appraisal and opinion dynamics take place on the same
timescale. Consider $n\geq 2$ individuals discussing a single issue on
timescale $k=0,1,2,\dots$ according to the FJ model
\begin{equation}\label{e12}
y(k+1)=\Theta W(x(k)) y(k)+(I_{n}-\Theta)y(0),
\end{equation}
where $W(x(k))=\diag(x(k))+(I_{n}-\diag(x(k)))C$, $x(k)$ is the
individuals' social power, $y(k)$ is the opinion vector, $\Theta$ is the
diagonal matrix describing individuals' susceptibilities to interpersonal
influence, and $C$ is the row-stochastic and zero-diagonal relative
interaction matrix. By equation (\ref{e12}), we have
\begin{equation}\label{e13}
y(k+1)=V(k+1)y(0),
\end{equation}
where $V(k+1)$ is row-stochastic, and satisfies $V(k+1)=\Theta W(x(k)) V(k)+I_{n}-\Theta$ with $V(0)=I_{n}$.

Similarly, in equation \eqref{e13}, the $i$-th column of $V(k+1)$ is the relative control of individual $i$'s initial opinion onto all others' opinions at time $k$. Based on the reflected appraisal mechanism, we suppose that each individual's self-appraisal at time $k+1$ equals its social power at time $k$, that is, $x(k+1)=V(k+1)^T\mathbf{1}_{n}/{n}$.

\begin{definition}\label{d2}
(The DeGroot-Friedkin model with stubborn individuals over a single issue) Consider an influence network with $n\geq 2$ individuals discussing a single issue over timescale $k=0,1,2,\dots$. Let $C$ and $\Theta=\diag(\theta_{1},\theta_{2},\dots,\theta_{n})$ be the row-stochastic, zero-diagonal relative interaction matrix and the diagonal matrix representing individuals' susceptibilities, respectively. Then, the DeGroot-Friedkin model with stubborn individuals over a single issue is
\begin{equation}\label{e14}
\begin{cases}
V(k+1)=\Theta W(x(k)) V(k)+I_{n}-\Theta,\\
x(k+1)=\dfrac{V(k+1)^T\mathbf{1}_{n}}{n},
\end{cases}
\end{equation}
with $W(x(k))=\diag(x(k))+(I_{n}-\diag(x(k)))C$ and $V(0)=I_{n}$.
\end{definition}

\begin{remark}\label{r2}
In the formulation of reflected appraisal mechanism \cite{NEF:11}, both individual's self-weights for current opinions and stubbornness are postulated as the reflected appraisals of its social power. In this paper, we focus on the case that individual's self-weights for its current opinions equal to its manifested social power. 
\end{remark}

Let $\Gamma_{n}=\setdef{W\in\mathbb{R}^{n\times n}}{W\geq 0, W\mathbf{1}_{n}=\mathbf{1}_{n}}$ denote the set of $n\times n$ row-stochastic real matrices. Define $\map{G}{\Gamma_{n}\times\Delta_{n}}{\Gamma_{n}\times\Delta_{n}}$ by $G(V,x)=(G_{V}(V,x),G_{x}(V,x))$ with $G_{V}(V,x)=\Theta W(x) V+I_{n}-\Theta$ and $G_{x}(V,x)=G_{V}(V,x)^T\mathbf{1}_{n}/n$. Then, system \eqref{e14} can be expressed by
\begin{equation*}
\begin{cases}
V(k+1)=G_{V}(V(k),x(k)),\\
x(k+1)=G_{x}(V(k),x(k)).
\end{cases}
\end{equation*}

\section{Equilibrium analysis}\label{s3}
This section studies the properties of the equilibria of models (\ref{e4})
and \eqref{e14}.

\subsection{Equivalence of equilibrium social power}\label{s3.1}
Since $F(x)$ and $G(V,x)$ are both continuous functions from, respectively, $\Delta_{n}$ and $\Gamma_{n}\times\Delta_{n}$ to themselves, where $\Delta_{n}$ and $\Gamma_{n}\times\Delta_{n}$ are convex and compact subsets of Banach space. Then, following the Schauder fixed point theorem \cite{VB:07}, i.e., every continuous function from a convex compact subset of a Banach space to itself has a fixed point, systems \eqref{e4} and \eqref{e14} have at least one equilibrium, respectively.
\begin{lemma}\label{L4}
(Equivalence of equilibrium social power) Suppose that Assumption~\ref{A1} holds, system \eqref{e4} and \eqref{e14} have the same relative interaction matrix $C$ and susceptibility matrix $\Theta$. Then, $x^{*}$ is an equilibrium of system \eqref{e4} if and only if for $V^{*}=(I_{n}-\Theta W(x^{*}))^{-1}(I_{n}-\Theta)\in\Gamma_{n}$, $(V^{*},x^{*})$ is an equilibrium of system \eqref{e14}.
\end{lemma}

Lemma \ref{L4} implies that the reflected appraisal mechanism is robust with respect to variations in the time scales at which opinions and social power evolve. Moreover, since non-stubborn individual has $0$ equilibrium social power in system \eqref{e4}, it also have $0$ equilibrium social power in system \eqref{e14}. Thus, we assume that Assumption~\ref{A2} holds for model (\ref{e14}) in the sequel.

\subsection{Properties of equilibrium social power with general topology}\label{s3.2}
Since systems \eqref{e4} and \eqref{e14} have same equilibrium social
power, we focus on equilibria of system \eqref{e4}.  In what follows, let
$\thetamin=\min_{j}\theta_{j}$, $\thetaave=\sum_{j=1}^{n}\theta_{j}/n$, and
$\thetamax=\max_{j}\theta_{j}$. Moreover, let $\Vfully$ and $\Vpartly$
denote the sets of individuals who are fully stubborn ($\theta_i=0$) and
partially stubborn ($\theta_i>0$), respectively. Without loss of
generality, assume $\Vfully=\{1,\dots,r\}$ and $\Vpartly=\{r+1,\dots,n\}$
with $r<n$.

\begin{lemma}\label{L4.1} 
(Properties of $F(x)$) For the map $\map{F}{\Delta_{n}}{\Delta_{n}}$ defined by $F(x)=(I_{n}-\Theta)(I_{n}-W(x)^{T}\Theta)^{-1}\mathbf{1}_{n}/{n}$ with $W(x)=\diag(x)+(I_{n}-\diag(x))C$, the following statements hold true:
\begin{enumerate}[itemindent=0pt]
\item $F$ is differentiable on $\interior\Delta_{n}$ and continuous on $\Delta_{n}$;
\item the Jacobian of $F$ is $\partial F/\partial x=(I_{n}-\Theta)(I_{n}-W(x)^{T}\Theta)^{-1}(I_{n}-C^{T})\Theta(I_{n}-\Theta)^{-1}\diag(F(x))$;
\item for any $x\in\Delta_{n}$, $(1-\theta_{i})/n\leq F_{i}(x) \leq(1+\zeta)/n$, where $\zeta=n\thetaave-\thetamin$.
\end{enumerate}
\end{lemma}

\begin{theorem}\label{T1}
(Equilibrium social power with general topology) Consider systems (\ref{e4}) and \eqref{e14} with $n\geq2$ and $x(0)\in\Delta_{n}$. Suppose that Assumption~\ref{A2} holds, and $C$ is row-stochastic and zero-diagonal. Then, we have that:
\begin{enumerate}
 \item there exists at least one equilibrium of systems \eqref{e4} and \eqref{e14}, which satisfies 
    \begin{enumerate}
    \item $x^{*}\in \interior\Delta_{n}$;
    \item $x^{*}_{i}\geq1/n$ for $i\in \Vfully$, and $x^{*}_{i}=1/n$ if and only if $C_{ji}=0$ for any $j\in \Vpartly$;
    \item $x^{*}_{i}>(1-\theta_{i})/n$ for $i\in \Vpartly$, and $x^{*}_{i}<1/n$ if $C_{ji}=0$ for any $j\in \Vpartly$;
    \item $\max_{i}x_{i}^{*}<1/n+\thetaave$.
    \end{enumerate}
    \item the equilibrium social power $x^{*}$ is unique if
      $\thetamax<\dfrac{n}{n+2(1+\zeta)}$ with $\zeta=n\thetaave-\thetamin$.
\end{enumerate}
\end{theorem}

\begin{remark}\label{r5}
In Theorem \ref{T1} we prove that if $\thetamax<\dfrac{n}{n+2(1+\zeta)}$, then $F(x)$ is contractive on $\Delta_{n}$, which also implies that the equilibrium social power only depends upon $C$ and $\Theta$. Since $\zeta<n-1$, then we have $\dfrac{n}{n+2(1+\zeta)}>1/3$, which implies that $\thetamax<\dfrac{n}{n+2(1+\zeta)}$ is a milder restriction compared with that proposed in \cite{AM-PJ-NEF-FB:13x}. Moreover, note that $\dfrac{n}{n+2(1+\zeta)}=\dfrac{1}{1+2\thetaave+\dfrac{2}{n}(1-\thetamin)}$, that is, $\thetamax<\dfrac{n}{n+2(1+\zeta)}$ is a restriction on the distribution of individuals' stubbornness. For clarification, now consider a special case. Suppose that  $r\geq1$. Then, we have that $\zeta<n-r$. Thus, it follows that $\dfrac{n}{n+2(1+\zeta)}>\dfrac{1}{1+2(1+1/n-r/n)}$, which tends to $\dfrac{1}{1+2/n}$ as $r/n$ approaches $1$. That is, $\thetamax$ can be arbitrarily close to $1$ in a large group where the majority is fully stubborn.
\end{remark}

Note that the relative interaction matrix $C$ is just required row-stochastic and zero-diagonal in Theorem \ref{T1}, which means that the the autocratic social power (i.e., there is exactly one individual has social power $1$, and all others' are $0$) can never emerge in systems \eqref{e4} and \eqref{e14}, even though the initial social power is autocratic or $\mathcal{G}(C)$ is star topology. This is a key difference between models \eqref{e4}, \eqref{e14} and the original DF model, in which the autocratic social power can be achieved under both irreducible and reducible influence networks \cite{PJ-AM-NEF-FB:13d}, \cite{PJ-NEF-FB:14m}. 

\begin{corollary}\label{C1}
(Properties of equilibrium social power) Consider systems (\ref{e4}) and \eqref{e14} with $n\geq2$ and $x(0)\in\Delta_{n}$. Suppose that Assumption~\ref{A2} holds, and $C$ is row-stochastic and zero-diagonal. Then the equilibrium social power of systems \eqref{e4} and \eqref{e14}, i.e., $x^{*}$, satisfies: 
 \begin{enumerate}
    \item for any $i\in \Vfully$ and $j\in \Vpartly$, if $C_{ki}=C_{kj}$ holds for any $k\in\Vpartly\setminus\{j\}$, then $x^{*}_{i}>x^{*}_{j}$;
    \item for any $i,j\in \Vpartly$, suppose that $C_{ki}=C_{kj}$ holds for any $k\in\Vpartly\setminus\{i,j\}$ and $C_{ij}=C_{ji}$. Then $x^{*}_{i}<x^{*}_{j}$ holds if and only if $\theta_{i}>\theta_{j}$;
    \item suppose that $C$ is symmetric. Then for any $i,j$, if $\theta_{i}>\theta_{j}$, then $x^{*}_{i}<x^{*}_{j}$.
    \end{enumerate}
\end{corollary}

Corollary \ref{C1} shows that if two individuals are accorded same influence weights by partially stubborn individuals, or the relative interaction matrix is symmetric, then the more stubborn individual has more equilibrium social power. In the DF model without stubbornness, the democratic social power structure, i.e., $x^{*}=\mathbf{1}_{n}/n$, is achieved only if the network is irreducible and doubly-stochastic. Next, we show that for systems (\ref{e4}) and \eqref{e14}, the democracy can be achieved even if the network is neither doubly-stochastic nor irreducible.

\begin{corollary}\label{C2}
(Existence of democratic equilibrium social power) Consider system (\ref{e4}) and \eqref{e14} with $n\geq2$ and $x(0)\in\Delta_{n}$. Suppose that Assumption~\ref{A2} holds, and $C$ is row-stochastic and zero-diagonal. Then, $\mathbf{1}_{n}/n$ is an equilibrium of systems (\ref{e4}) and \eqref{e14} if and only if $\Theta(I_{n}-\Theta)^{-1}\mathbf{1}_{n}$ is a left eigenvector of $C$ corresponding to eigenvalue $1$.
\end{corollary}
The proof of Corollary \ref{C2} can be readily obtained by substituting $x$ and $F(x)$ for $x^{*}=\mathbf{1}_{n}/n$ in equation (\ref{e4}). 

\subsection{Properties of equilibrium social power with star topology}\label{s3.3}

First, we consider the scenario where the center node of $\mathcal{G}(C)$ belongs to $\Vfully$. 

\begin{theorem}\label{T2}
(Equilibrium social power under star topology with fully stubborn center node) Consider system (\ref{e4}) and \eqref{e14} with $n\geq2$ and $x(0)\in\Delta_{n}$. Suppose that Assumption~\ref{A2} holds, and $C$ is row-stochastic and zero-diagonal with $\mathcal{G}(C)$ being a star topology with center node $l$ satisfying $\theta_{l}=0$. Then, the equilibrium social power of systems (\ref{e4}) and \eqref{e14} is unique, and satisfies:
 \begin{enumerate}
\item $x^{*}\in \interior\Delta_{n}$;
\item $x^{*}_{i}=1/n$ for $i\in \Vfully\setminus\{l\}$;
\item $x^{*}_{i}=\dfrac{n-\sqrt{n^{2}-4n\theta_{i}(1-\theta_{i})}}{2n\theta_{i}}<\dfrac{1}{n}$, and decreases with respect to $\theta_{i}$ for $i\in\Vpartly$;
\item $x^{*}_{l}=\dfrac{1}{n}+\dfrac{1}{n}\sum\limits_{j=r+1}^{n}\dfrac{\theta_{j}(1-x^{*}_{j})}{1-\theta_{j}x^{*}_{j}}>\dfrac{1}{n}$.
 \end{enumerate} 
\end{theorem}

Theorem \ref{T2} shows that for systems (\ref{e4}) and \eqref{e14} under star topology with fully stubborn center node, the center node has the largest equilibrium social power, which is strictly larger than $1/n$. And other fully stubborn individuals' equilibrium social power is $1/n$, while all partially stubborn individuals' equilibrium social power is strictly less than $1/n$. Moreover, the ordering of equilibrium social power of partially stubborn individuals is consistent with the ordering of their stubbornness. Now, we consider the scenario where the center node of $\mathcal{G}(C)$ belongs to $\Vpartly$. 

\begin{theorem}\label{T3}
(Equilibrium social power under star topology with partially stubborn center node) Consider systems (\ref{e4}) and \eqref{e14} with $n\geq2$ and $x(0)\in\Delta_{n}$. Suppose that Assumption~\ref{A2} holds, and $C$ is row-stochastic and zero-diagonal with $\mathcal{G}(C)$ being a star topology with center node $l$ satisfying $1>\theta_{l}>0$. Then, 
\begin{enumerate}
\item the equilibrium social power of systems (\ref{e4}) and \eqref{e14} has the following properties:
\begin{enumerate}
\item $x^{*}\in \interior\Delta_{n}$;
\item for $i\in \Vfully$, if $C_{li}=0$, then $x^{*}_{i}=1/n$; otherwise, $x^{*}_{i}>1/n$;
\item  for $i\in \Vpartly\setminus\{l\}$, if $C_{li}=0$, then $x^{*}_{i}$ is unique, $x^{*}_{i}=\dfrac{n-\sqrt{n^{2}-4n\theta_{i}(1-\theta_{i})}}{2n\theta_{i}}$, and decreases with respect to $\theta_{i}$.
\end{enumerate}
\item Moreover, if there holds $C_{li}=0$ for all $i\in \Vpartly\setminus\{l\}$, then the equilibrium social power of systems (\ref{e4}) and \eqref{e14} is unique, and satisfies:
\begin{enumerate}
\item  $x^{*}_{i}=\dfrac{n-\sqrt{n^{2}-4n\theta_{i}(1-\theta_{i})}}{2n\theta_{i}}$, and decreases with respect to $\theta_{i}$ for $i\in \Vpartly\setminus\{l\}$;
\item  $x^{*}_{l}=\dfrac{n-\sqrt{n^{2}-4n\theta_{l}(1-\theta_{l})\xi^{*}}}{2n\theta_{l}}$;
\item  $x^{*}_{i}=\dfrac{1}{n}+(\dfrac{\xi^{*}}{n}-x^{*}_{l})C_{li}$ for $i\in \Vfully$, 
\end{enumerate}
where $\xi^{*}=n-r-n\sum_{j\in \Vpartly\setminus\{l\}}x^{*}_{j}$.
\end{enumerate}
\end{theorem}

Theorem \ref{T3} shows that all individuals have positive equilibrium social power, while the partially stubborn center does not necessarily have the largest equilibrium social power. The following examples show that under the same star topology with partially stubborn center node, both fully stubborn individual and partially stubborn individual (whether if it is center node or not) can obtain the largest equilibrium social power, which depends upon individuals' stubbornness.

\begin{figure}
 \centering
  \includegraphics[width=\hsize]{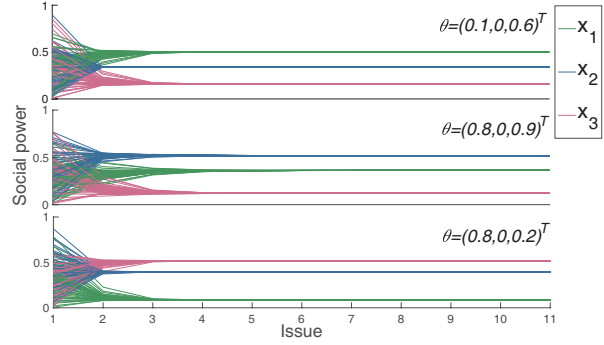}\\
  \caption{$50$ runs of trajectories of $x(s)$ under different $\Theta$.}\label{fig1}
\end{figure}

\paragraph*{Numerical examples on star topology with partially stubborn center node}
Consider system \eqref{e4} with $n=3$. Suppose that $C=[0, 0.2, 0.8; 1, 0, 0; 1, 0, 0]$, i.e., individual $1$ is the center node. Then, under different settings of $\Theta$, we obtain the trajectories of $x(s)$, shown in Fig. (\ref{fig1}).
It is observed that in Fig. (\ref{fig1}), the center node $1$ occupies the largest equilibrium social power when $\theta=(0.1, 0, 0.6)^{T}$, while the fully stubborn node and partially stubborn node which are not center node can also obtain largest equilibrium social power under the same influence network but different settings of $\theta$.

\begin{corollary}\label{C3}
(Ordering of equilibrium social power under star topology with partially stubborn center node) Consider systems (\ref{e4}) and \eqref{e14} with $n\geq2$ and $x(0)\in\Delta_{n}$. Suppose that Assumption~\ref{A2} holds, and $C$ is row-stochastic and zero-diagonal with $\mathcal{G}(C)$ being a star topology with center node $l$ satisfying $1>\theta_{l}>0$. Then, the equilibrium social power of systems (\ref{e4}) and \eqref{e14} satisfies:
\begin{enumerate}
\item for any $i,j\in \Vfully$, if $C_{li}>C_{lj}$, then $x^{*}_{i}>x^{*}_{j}$;
\item  for any $i\in \Vfully$ and $j\in \Vpartly\setminus\{l\}$, if $C_{li}=C_{lj}$, then $x^{*}_{i}>x^{*}_{j}$;
\item  for any $i,j\in \Vpartly\setminus\{l\}$ with $C_{li}=C_{lj}$, $x^{*}_{i}>x^{*}_{j}$ if and only if $\theta_{i}<\theta_{j}$;
\item  for any $i,j\in \Vpartly\setminus\{l\}$ with $\theta_{i}=\theta_{j}$, $x^{*}_{i}>x^{*}_{j}$ if and only if $C_{li}>C_{lj}$.
\end{enumerate}
\end{corollary}

\section{Convergence analysis}\label{s4}

This section studies the convergence of systems \eqref{e4} and \eqref{e14}. 

\subsection{Convergence of the DF model with stubborn individuals over issue sequences}\label{s4.1}

\begin{theorem}\label{T4}
(Convergence with general topology) Consider system (\ref{e4}) with
  $n\geq2$ and $x(0)\in\Delta_{n}$. Suppose that Assumption~\ref{A2} holds,
  and $C$ is row-stochastic and zero-diagonal. Let
  $\zeta=n\thetaave-\thetamin$. If $\thetamax<\dfrac{n}{n+2(1+\zeta)}$,
  then all trajectories of system (\ref{e4}) converge to the unique
  equilibrium social power $x^{*}$ characterized in Theorem \ref{T1}
  exponentially fast.
\end{theorem}

In the proof of Theorem \ref{T1}, we show that if $\thetamax<\dfrac{n}{n+2(1+\zeta)}$, $F(x)$ is contractive on $\Delta_{n}$. Then the exponential convergence of system (\ref{e4}) follows from the Banach fixed point theorem. Next, we consider the convergence of system (\ref{e4}) with star topology. By the proof of Theorem \ref{T2}, we have the following Corollary.

\begin{corollary}\label{C4}
(Convergence under star topology with fully stubborn center node) Consider system (\ref{e4}) with $n\geq2$ and $x(0)\in\Delta_{n}$. Suppose that Assumption~\ref{A2} holds, and $C$ is row-stochastic and zero-diagonal with $\mathcal{G}(C)$ being a star topology with center node $l$ satisfying $\theta_{l}=0$. Then, all trajectories of system (\ref{e4}) exponentially converge to the unique equilibrium social power $x^{*}$ characterized in Theorem \ref{T2}.
\end{corollary}

Next we consider the case that the center node is partially stubborn.

\begin{theorem}\label{T6}
(Convergence property under star topology with partially stubborn center node) Consider system (\ref{e4}) with $n\geq2$ and $x(0)\in\Delta_{n}$. Suppose that Assumption~\ref{A2} holds, and $C$ is row-stochastic and zero-diagonal with $\mathcal{G}(C)$ being a star topology with center node $l$ satisfying $1>\theta_{l}>0$. Then, 
\begin{enumerate}
\item for $i\in \Vpartly\setminus\{l\}$, if $C_{li}=0$, then $x_{i}(s)$ exponentially converges to $x^{*}_{i}=\dfrac{n-\sqrt{n^{2}-4n\theta_{i}(1-\theta_{i})}}{2n\theta_{i}}$;
\item moreover, if there holds $C_{li}=0$ for all $i\in \Vpartly\setminus\{l\}$ and $\sum_{j\in \Vpartly/\{l\}}\theta_{j}\leq 4n/5-1$, then all trajectories of system (\ref{e4}) exponentially converge to the equilibrium social power $x^{*}$ characterized in statement (ii) of Theorem \ref{T3}.
\end{enumerate}
\end{theorem}

\subsection{Convergence of the DF model with stubborn individuals over a single issue}\label{s4.2}

First, we consider doubly-stochastic influence network with uniformly stubborn individuals.

\begin{lemma}\label{L7}
(Convergence with doubly-stochastic topology and uniform stubbornness) Consider system (\ref{e14}) with $n\geq2$ and $x(0)\in\Delta_{n}$. Suppose that $\theta_{i}=\theta$ with $\theta\in(0,1)$ for all $i\in\{1,2,...,n\}$, $C$ is doubly-stochastic and zero-diagonal. Then all trajectories of system (\ref{e14}) exponentially converge to the democratic equilibrium $\mathbf{1}_{n}/n$.
\end{lemma}

Since $\theta_{i}=\theta$ and $C$ is doubly-stochastic, we have $x(k+1)=\theta x(k)+(1-\theta)\mathbf{1}_{n}/n$. Note that $\theta\in(0,1)$, thus $x(k)\to \mathbf{1}_{n}/n$. It is clear that in system (\ref{e14}), if $V(k)$ converges, then $x(k)$ converges. Let $V\in\Gamma_{n}$ be a row-stochastic matrix, and $V_{i}$ denote the $i$-th column of $V$. Let $\chi=[V_{1}^{T} \ V_{2}^{T}\ ...\ V_{n}^{T} ]^{T}$ denote the vector by vectorizing $V$, then $\chi\in \mathcal{A}=\{x\mid x\in\mathbb{R}^{n^{2}},x\geq 0, \sum_{i=0}^{n-1} x_{ni+t}=1 \ {\rm for \ any} \ t\in\{1,2,...,n\}\}$. Let $\nu=[(1-\theta_{1})\mathbf{e}_{1}^{T} (1-\theta_{2})\mathbf{e}_{2}^{T} ... (1-\theta_{n})\mathbf{e}_{n}^{T}]^{T}\in\mathbb{R}^{n^{2}}$. Define $\map{\hat{G}}{\mathcal{A}}{\mathcal{A}}$ by
\begin{equation*}
\hat{G}(x)=I_{n}\otimes \Theta W(x)x+\nu,
\end{equation*}
where $x\in\mathcal{A}$, $W(x)=\diag(\omega)+(I_{n}-\diag(\omega))C$ with $\omega\in\mathbb{R}^{n}$ and $\omega_{i}=V_{i}^{T}\mathbf{1}_{n}/n$. Now, we present our convergence result for system (\ref{e14}) with general topology.

\begin{theorem}\label{T8}
(Convergence with general topology) Consider system (\ref{e14}) with $n\geq2$ and $x(0)\in\Delta_{n}$. Suppose that Assumption~\ref{A2} holds, and $C$ is row-stochastic and zero-diagonal. If $\thetamax<1/2$, then, all trajectories of system (\ref{e14}) converge to the unique equilibrium social power $x^{*}$ characterized in Theorem \ref{T1} exponentially fast.
\end{theorem}
Note that even though systems \eqref{e4} and \eqref{e14} have the same equilibrium social power, their trajectories may be different. In the proof of Theorem \ref{T8}, we show that system (\ref{e14}) is contractive if $\thetamax<1/2$. However, this condition is not necessary. In next section, we will propose a conjecture on the contractivity of systems \eqref{e4} and \eqref{e14}.
\section{Simulations and conjecture}\label{s5}

As we shown, the equilibrium social power of systems \eqref{e4} and \eqref{e14} is unique if $\thetamax<\dfrac{n}{n+2(1+\zeta)}$. However, for the general case, the uniqueness of equilibrium social power of systems \eqref{e4} and \eqref{e14} is equivalent to that the quadratic equations $(I_{n}-C^{T}\Theta)(I_{n}-\Theta)^{-1}x-(I_{n}-C^{T})\Theta(I_{n}-\Theta)^{-1}\diag(x)x=\mathbf{1}_{n}/n$ has exactly one solution in $\interior\Delta_{n}$, which is difficult to prove due to the entanglement of $C$ and $\Theta$. In this subsection, we shall estimate the probability that systems \eqref{e4} and \eqref{e14} converge to unique equilibrium social power for any initial social power and matrix pair $(C,\Theta)$.  

\paragraph*{Monte Carlo validation} 
Since systems \eqref{e4} and \eqref{e14} have the same equilibrium social power, here we just focus on system \eqref{e4}. For given matrix pair $(C,\Theta)$, where $C$ is row-stochastic and zero-diagonal, $\Theta$ satisfies Assumption~\ref{A2}, we randomly pick $\hat{x}(0)$ and compute the final social power $\hat{x}^{*}$ by running system \eqref{e4}. Let $x\in \Delta_{n}$ be a random variable representing the initial social power, and $x^{*}_{x}$ denote the corresponding final social power of system \eqref{e4}. Then, define $pr(C,\Theta)=Pr\{J(x)=0\}$ as the probability that system \eqref{e4} converges to $\hat{x}^{*}$ with initial social power $x$, where $\map{J}{\Delta_{n}}{\mathbb{R}^{n}}=x^{*}_{x}-\hat{x}^{*}$ is a measurable performance function. Now, we can estimate $pr(C,\Theta)$ as follows. First, we generate $N$ independent identically distributed random samples of the initial social power $x^{1},x^{2},\dots,x^{N}$, where $N$ is a positive integer. Second, define an indicator function $\map{\mathbb{I}_{J,C,\Theta}}{\Delta_{n}}{\{0,1\}}$ by $\mathbb{I}_{J,C,\Theta}(x)=1$ if $J(x)=0$, and $0$ otherwise. Finally, we compute the empirical probability as   
\begin{equation*}
\hat{pr}(C,\Theta)=\dfrac{1}{N}\sum_{i=1}^{N}\mathbb{I}_{J,C,\Theta}(x^{i}).
\end{equation*}
Then, for any accuracy $\epsilon\in(0,1)$ and confidence level $1-\eta\in(0,1)$, by the Chernoff bound we have that
\begin{equation*}
\begin{split}
Pr\{\mid\hat{pr}(C,\Theta)-pr(C,\Theta)\mid<\epsilon\}\geq 1-2\exp(-2\epsilon^{2}N).
\end{split}
\end{equation*}
If there holds $N\geq\log(2/\eta)/(2\epsilon^{2})$, then we have $1-2\exp(-2\epsilon^{2}N)>1-\eta$, that is, the probability that $\mid\hat{pr}(C,\Theta)-pr(C,\Theta)\mid<\epsilon$ is greater than $1-\eta$. In \cite{GN-FB:07z}, the authors computed that for $\epsilon=\eta=0.01$, the Chernoff bound is satisfied by $N=27000$. That is to say, for given $(C,\Theta)$, if system \eqref{e4} converges to $x^{*}_{\mathbf{e}_{1}}$ for all $27000$ samples of initial social power, we can say that for the given $C$ and $\Theta$, with confidence level $99\%$, there is at least $99\%$ probability that system \eqref{e4} converges to unique equilibrium social power for any initial social power. 

Similarly, consider random variable $(C,\Theta)$ where $C$ is row-stochastic and zero-diagonal, $\theta_{i}\in[0,1)$ for $i\in\until n$. If there holds that for each of $27000$ samples of matrix pairs, system \eqref{e4} converges to same equilibrium social power for all $27000$ samples of initial social power, then we can say that for any $C$ and $\Theta$ satisfying Assumption~\ref{A2}, with confidence level $99\%$, there is at least $99\%$ probability that system \eqref{e4} converges to unique equilibrium social power for any initial social power. 

\begin{figure}
\centering
 {
\begin{minipage}[t]{0.25\textwidth}
 \centering
  \includegraphics[width=\hsize]{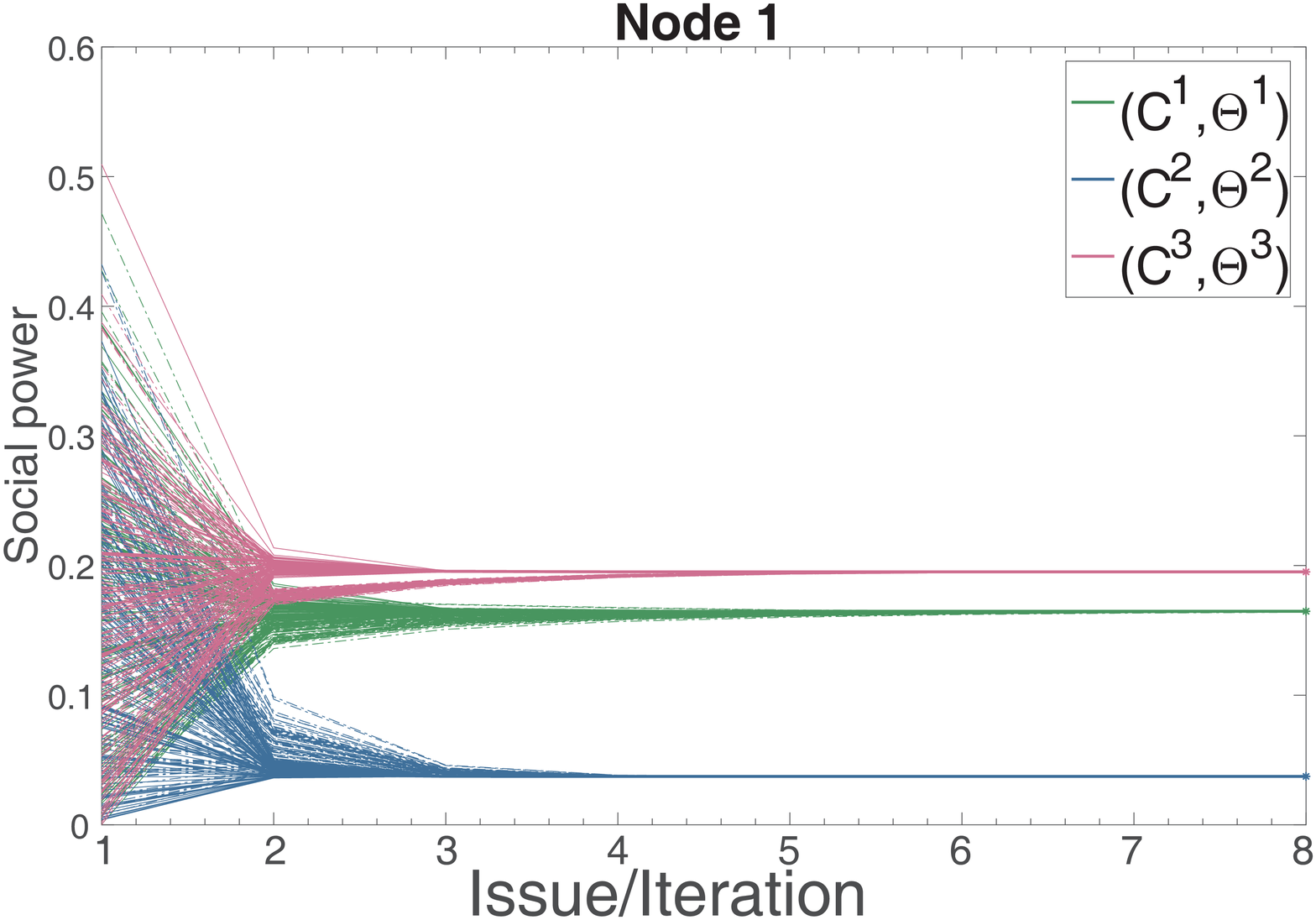}\\
 \end{minipage}
}
\hspace{-5.5ex}
{
\begin{minipage}[t]{0.25\textwidth}
 \centering
  \includegraphics[width=\hsize]{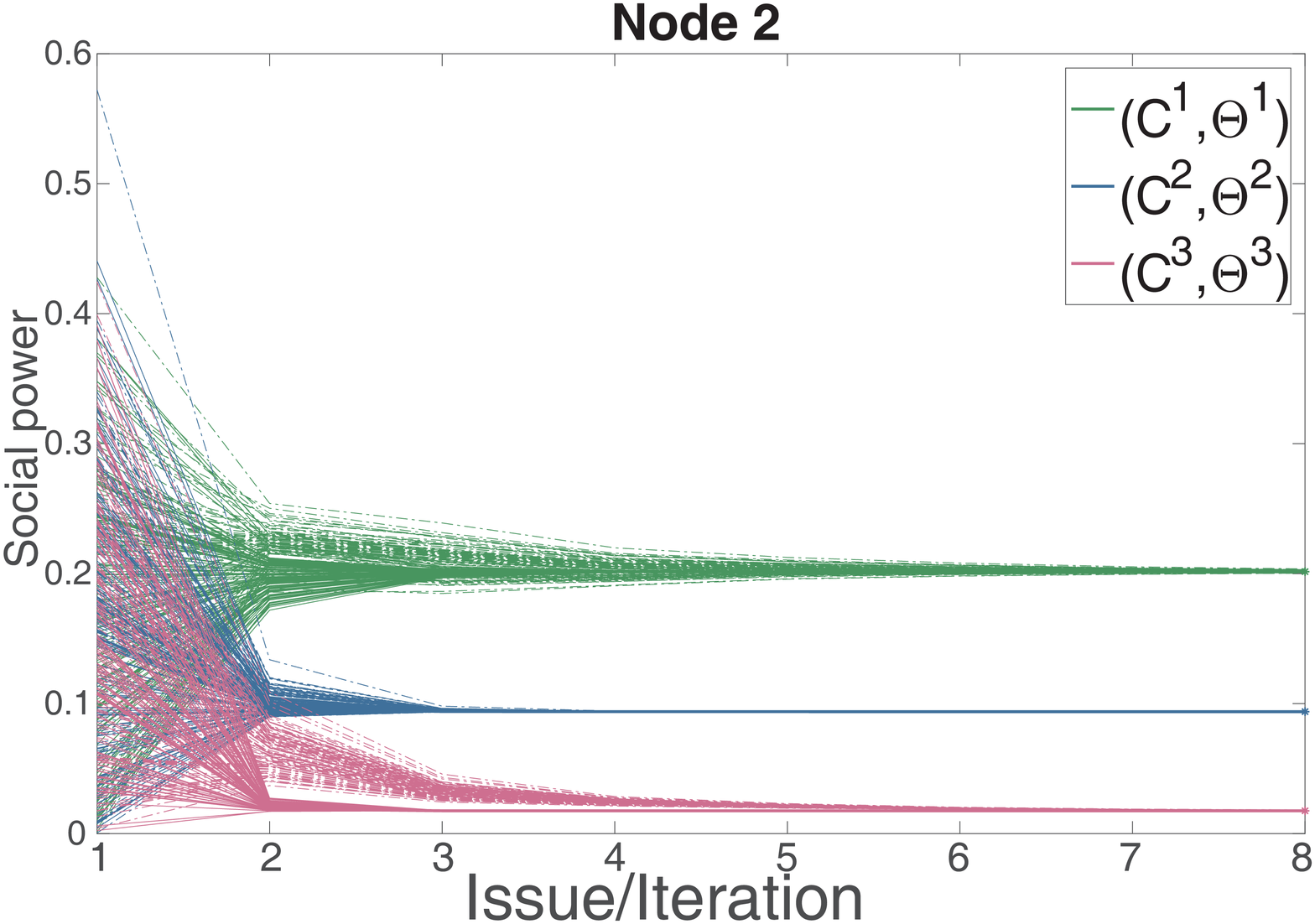}\\
  \end{minipage}
}

{
\begin{minipage}[t]{0.25\textwidth}
 \centering
  \includegraphics[width=\hsize]{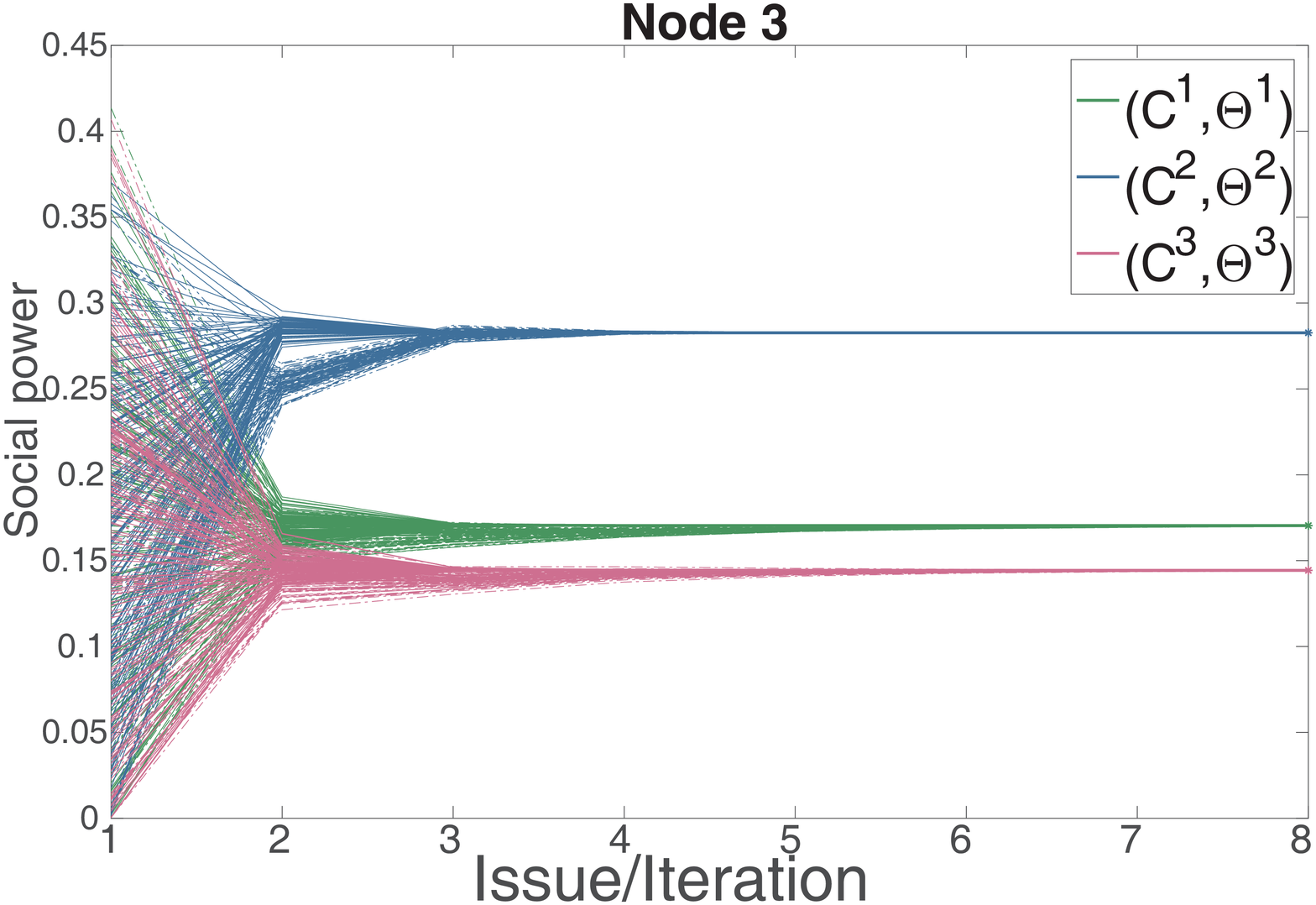}\\
  \end{minipage}
}
\hspace{-5.5ex}
{
\begin{minipage}[t]{0.25\textwidth}
 \centering
  \includegraphics[width=\hsize]{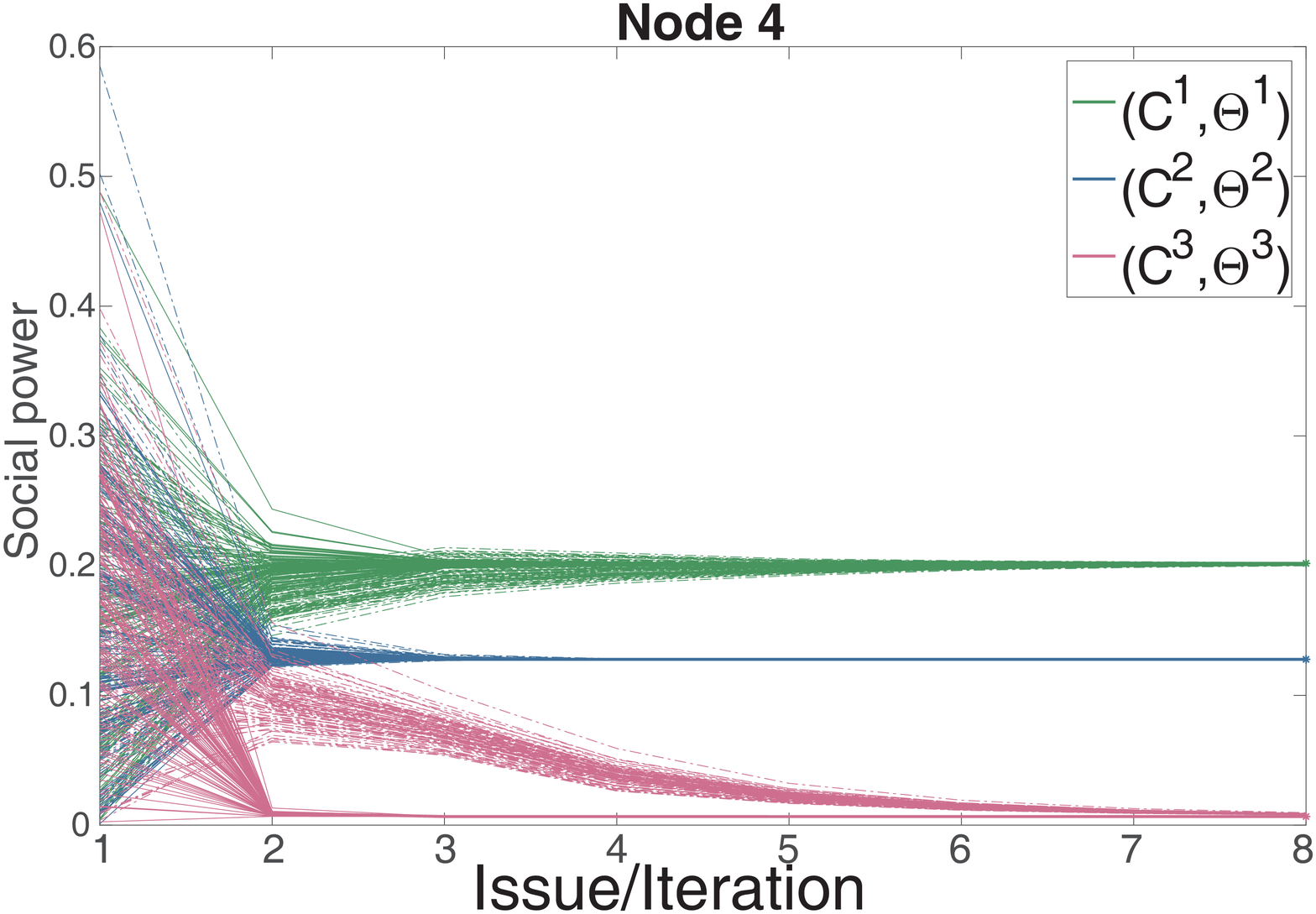}\\
  \end{minipage}
}

{
\begin{minipage}[t]{0.25\textwidth}
 \centering
  \includegraphics[width=\hsize]{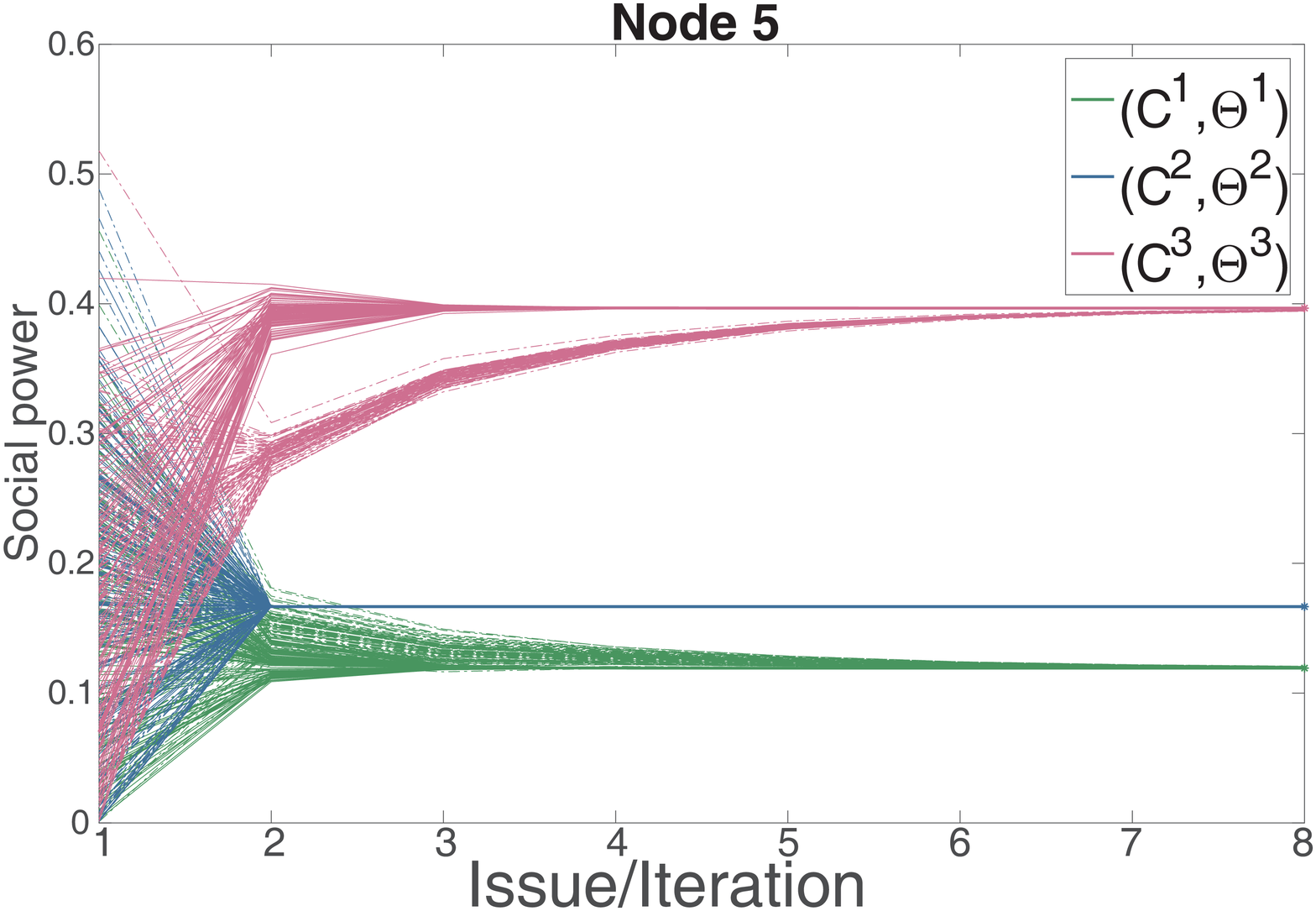}\\
  \end{minipage}
}
\hspace{-5.5ex}
{
\begin{minipage}[t]{0.25\textwidth}
 \centering
  \includegraphics[width=\hsize]{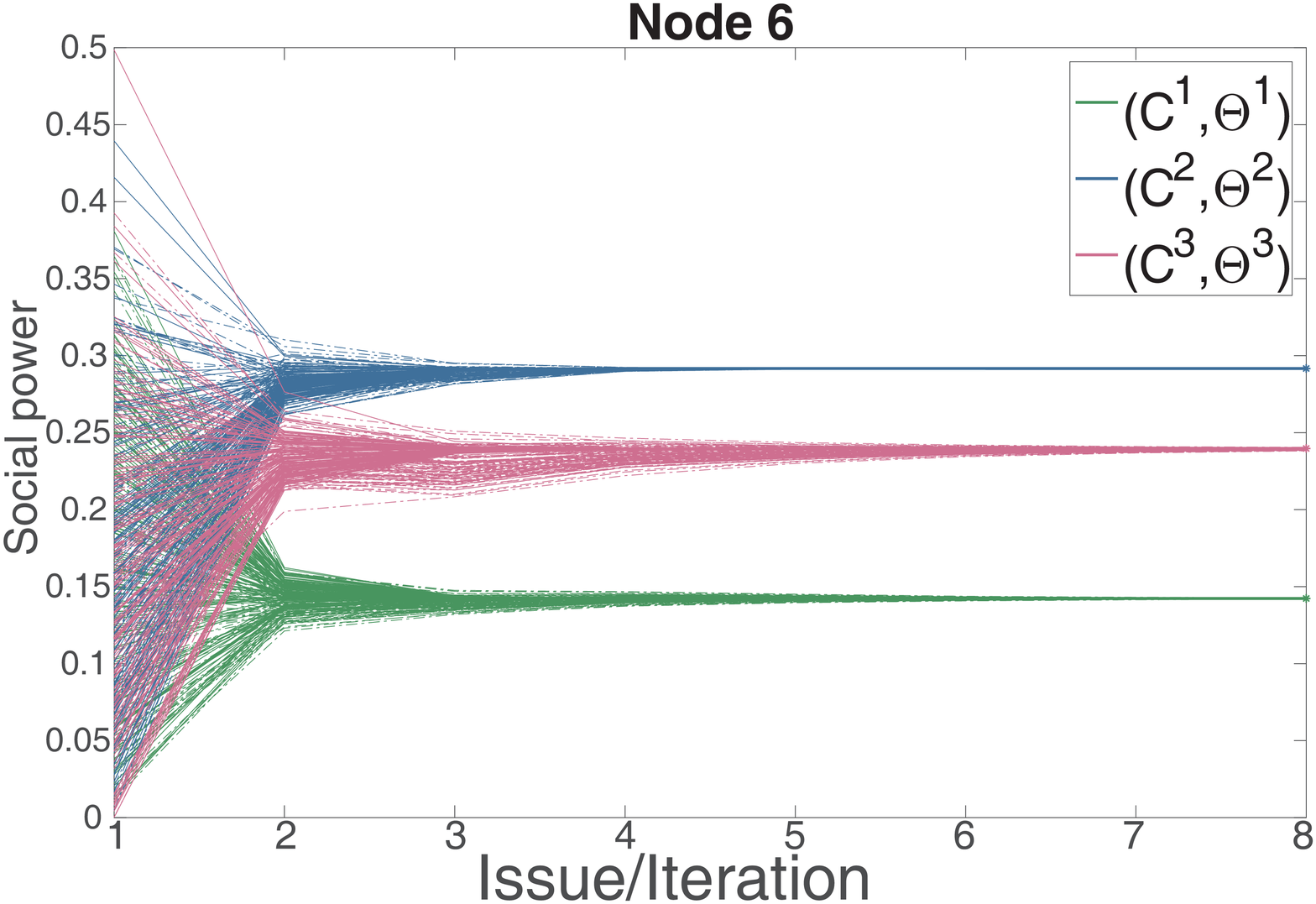}\\
  \end{minipage}
}
\hspace{0pt}
\caption{The trajectories for $6$ nodes of systems \eqref{e4} and \eqref{e14} beginning at $100$ samples of initial social power with $3$ samples of matrix pairs $(C^{i},\Theta^{i})$. The solid lines and dot lines dipict the trajectories of system \eqref{e4} and \eqref{e14}, respectively.}\label{fig3}
\end{figure}

\paragraph*{Numerical examples on uniqueness and convergence}
Based on above discussion, we run $27000^{2}$ experiments for systems \eqref{e4} and \eqref{e14} with randomly generated initial social power $x^{i}$, $i\in\until{27000}$ for each randomly generated matrix pair $(C^{j},\Theta^{j})$, $j\in\until{27000}$. Figure \ref{fig3} depicts the trajectories of $6$ nodes for $100$ initial social power with $3$ matrix pairs. The experiments show that for systems \eqref{e4} and \eqref{e14} with each of the $27000$ samples of $(C,\Theta)$, the trajectories beginning at all $27000$ samples of initial social power converge to the same equilibrium social power. Therefore, our experiments establish the following conjecture.

\begin{conjecture}\label{con1}
Consider system (\ref{e4}) and \eqref{e14} with $n\geq2$ and $x(0)\in\Delta_{n}$. Suppose that Assumption~\ref{A2} holds, and $C$ is row-stochastic and zero-diagonal. Then, all trajectories of systems (\ref{e4}) and \eqref{e14} exponentially converge to an unique equilibrium social power, which only depends upon $C$ and $\Theta$.
\end{conjecture}

\section{Conclusions}\label{s6}
This paper has investigated the evolution of social power of stubborn individuals. Two models are proposed to characterize the social power evolution over issue sequences and over a single issue, respectively. Analytical and numerical results are provided. We prove that the model over a single issue has the same equilibrium social power with the model over issue sequences. Based on this equivalence, uniqueness and properties of the equilibrium social power are analyzed under different settings of the influence network topology. Then, we establish convergence of the equilibrium.

Our investigations reveal several features for social power evolution of stubborn individuals. First, the reflected appraisal mechanism is robust with respect to variations in the time scales at which opinions and social power evolve. Second, individuals will exponentially forget their initial social power, and the equilibrium social power only depends upon the relative interaction matrix and their stubbornness. Third, individuals will have positive equilibrium social power if they are stubborn, and more stubbornness leads to more social power. Finally, for an influence network in which all individuals are stubborn, the autocratic social power structure never emerges, while the democratic social power can be achieved with any network topologies. Future works will focus on the co-evolution of individuals' stubbornness with their social power.

\bibliographystyle{plainurl}
\bibliography{alias,Main,FB,New}
\appendices

\section{Proof of Lemma \ref{L4}}
Regarding necessity, suppose that $x^{*}\in\Delta_{n}$ is an equilibrium of system (\ref{e4}). Then, by equation (\ref{e4}) we have $x^{*}=(I_{n}-\Theta)(I_{n}-W(x^{*})^{T}\Theta)^{-1}\mathbf{1}_{n}/n$. Let $V^{*}=(I_{n}-\Theta W(x^{*}))^{-1}(I_{n}-\Theta)$, then $V^{*}\in\Gamma$ and $x^{*}=(V^{*})^{T}\mathbf{1}_{n}/n$. Since
\begin{equation*}
\begin{split}
&\Theta W(x^{*})V^{*}+I_{n}-\Theta\\
=&(\Theta W(x^{*})\!+\!I_{n}\!-\!\Theta W(x^{*}))(I_{n}-\Theta W(x^{*}))^{-1}(I_{n}\!-\!\Theta)\\
=&V^{*},
\end{split}
\end{equation*}
then $(V^{*},x^{*})$ is an equilibrium of system \eqref{e14}.

Regarding sufficiency, suppose that $(V^{*},x^{*})$ is an equilibrium of system \eqref{e14}. By equation \eqref{e14} we have $V^{*}=\Theta W(x^{*})V^{*}+I_{n}-\Theta$ and $x^{*}=(V^{*})^{T}\mathbf{1}_{n}/n$. Note that under Assumption~\ref{A1}, $I-\Theta W(x^{*})$ is nonsingular, thus $V^{*}=(I_{n}-\Theta W(x^{*}))^{-1}(I_{n}-\Theta)$. Then, $x^{*}=(I_{n}-\Theta)(I_{n}-W(x^{*})^{T}\Theta)^{-1}\mathbf{1}_{n}/n$, which implies that $x^{*}$ is an equilibrium of system \eqref{e4}. 

\section{Proof of Lemma \ref{L4.1}}
Denote $\Lambda(x)=I_{n}-W(x)^{T}\Theta$. Regarding (i), note $(I_{n}-W(x)^{T}\Theta)^{-1}=\Lambda^{*}(x)/\det(\Lambda(x))$, where $\Lambda^{*}(x)$ is the adjoint matrix of $\Lambda(x)$. Then,
\begin{equation*}
F_{i}(x)=\dfrac{(1-\theta_{i})}{n\times \det(\Lambda(x))}\sum_{k=1}^{n}\Lambda^{*}_{ik}(x),
\end{equation*}
where $\det(\Lambda(x))$ and $\Lambda^{*}_{ik}(x)$ are both analytic functions of $x$. Since $\det(\Lambda(x))\neq 0$, $F_{i}(x)$ is differentiable on $\interior\Delta_{n}$ and continuous on $\Delta_{n}$ for any $i$. That is, $F(x)$ is differentiable on $\interior\Delta_{n}$ and continuous on $\Delta_{n}$.

Regarding (ii), since $\theta_{i}<1$, $(I_{n}-\Theta)^{-1}$ exists. By \eqref{e7} we obtain $(I_{n}-W(x)^{T}\Theta)(I_{n}-\Theta)^{-1}F(x)=\mathbf{1}_{n}/n$. Then, taking the derivatives of both sides, we get
\begin{multline*}
(I_{n}-W(x)^{T}\Theta)(I_{n}-\Theta)^{-1}\dfrac{\partial F}{\partial x}\\
=(I_{n}-C^{T})\Theta(I_{n}-\Theta)^{-1}\diag(F(x)).
\end{multline*}
Hence, $\partial F/\partial x=(I_{n}-\Theta)(I_{n}-W(x)^{T}\Theta)^{-1}(I_{n}-C^{T})\Theta(I_{n}-\Theta)^{-1}\diag(F(x))$.

Regarding (iii), since $\Lambda^{-1}(x)(I_{n}-W(x)^{T}\Theta)=I_{n}$, we have that
\begin{equation*}
\Lambda^{-1}_{ii}(x)(1-\theta_{i}x_{i})-\theta_{i}(1-x_{i})\sum_{k=1}^{n}\Lambda^{-1}_{ik}(x)C_{ik}=1.
\end{equation*}
For any $i\in\until n$, since $1-\theta_{i}x_{i}>0$, there holds
\begin{equation*}
\Lambda^{-1}_{ii}(x)\!=\!\dfrac{1+\theta_{i}(1-x_{i})\sum_{k=1}^{n}\Lambda^{-1}_{ik}(x)C_{ik}}{1-\theta_{i}x_{i}}\geq\!1.
\end{equation*}
By $F(x)=(I_{n}-\Theta)\Lambda^{-1}(x)\mathbf{1}_{n}/n$, we have that
\begin{equation*}
F_{i}(x)\geq \dfrac{1}{n}(1-\theta_{i})\Lambda^{-1}_{ii}(x)\geq \dfrac{1-\theta_{i}}{n}>0.
\end{equation*}
On the other hand, by $\sum_{k=1}^{n}F_{k}(x)=1$ there holds
\begin{equation*}
F_{i}(x)=1-\sum_{k\neq i}F_{k}(x)\leq 1-\sum_{k\neq i}\dfrac{1-\theta_{k}}{n}=\dfrac{1}{n}+\dfrac{1}{n}\sum_{k\neq i}\theta_{k}.
\end{equation*}
Since $\sum_{k\neq i}\theta_{k}\leq n\thetaave-\thetamin=\zeta$ for any $i\in\until n$, we obtain $F_{i}(x)\leq(1+\zeta)/n$. 

\section{Proof of Theorem \ref{T1}}
Since the equilibrium social power of systems (\ref{e4}) and (\ref{e14}) is equivalent, we just need to show that (i) and (ii) hold for system (\ref{e4}). Regarding (i), by Lemma \ref{L4.1} we have that $F_{i}(x)\in (0,1)$, i.e., $x^{*}\in \interior\Delta_{n}$. According to the definitions of $\Vfully$ and $\Vpartly$, we write
\begin{equation*}
\Theta=\begin{bmatrix}
 0_{r\times r} & 0 \\
  0 & \ \ \Theta_{p}
  \end{bmatrix}, \ \ {\rm and} \ \ C=\begin{bmatrix}
 C_{f} & C_{fp} \\
  C_{pf} & \ \ C_{p}
  \end{bmatrix}.
\end{equation*}
Let $x_{f}^{*}\in\mathbb{R}^{r}$ and $x_{p}^{*}\in\mathbb{R}^{n-r}$ denote the equilibrium social power vectors of fully stubborn and partially stubborn individuals, respectively. Then, by equation (\ref{e4}),
\begin{equation}\label{e11}
\begin{cases}
x_{f}^{*}\!=\!\dfrac{\mathbf{1}_{r}}{n}\!+\!C_{pf}^{T}\!\Theta_{p}\!(\!I_{n-r}\!-\!\Theta_{p}\!)^{-1}\!(\!I_{n-r}\!-\!\diag(\!x_{p}^{*}\!)\!)x_{p}^{*},\\
\begin{split}
&\!(I_{n-r}-C_{p}^{T}\!\Theta_{p})(\!I_{n-r}\!-\!\Theta_{p}\!)^{-1}x_{p}^{*}=\dfrac{\mathbf{1}_{n-r}}{n}+\\
&\!(\!I_{n-r}\!-\!C_{p}^{T}\!)\Theta_{p}(I_{n-r}\!-\!\Theta_{p})^{-1}\!\diag(x_{p}^{*})x_{p}^{*}.
\end{split}
\end{cases}
\end{equation}
Since $x^{*}\in \interior\Delta_{n}$, we have $\diag(x_{p}^{*})x_{p}^{*}>0$ and $(I_{n-r}-\diag(x_{p}^{*}))x_{p}^{*}>0$, which imply that $x^{*}_{i}\geq1/n$ and $x^{*}_{j}\geq(1-\theta_{j})/n+\theta_{j}(x^{*}_{j})^{2}$ for any $i\in \mathcal{V}_{f}, j\in\mathcal{V}_{p}$. Moreover, for $i\in \mathcal{V}_{f}$, if $C_{ji}=0$ for all $j\in\mathcal{V}_{p}$, we have that $x^{*}_{i}=1/n$; otherwise, $x^{*}_{i}>1/n$. For $i\in\mathcal{V}_{p}$, if $C_{ji}=0$ for all $j\in\mathcal{V}_{p}$, we have that $x^{*}_{i}=(1-\theta_{i})/n+\theta_{i}(x^{*}_{i})^{2}$. Since $x^{*}_{i}<1$ and $n\geq 2$, there holds $x^{*}_{i}=\dfrac{n-\sqrt{n^{2}-4n\theta_{i}(1-\theta_{i})}}{2n\theta_{i}}<\dfrac{1}{n}$. Finally, $\max_{i}x_{i}^{*}<1/n+\thetaave$ follows from that $x^{*}\in\Delta_{n}$.

Regarding (ii), first, we show that $F(x)$ is contractive on $\Delta_{n}$ if $\thetamax<\dfrac{n}{n+2(1+\zeta)}$. Since $\norm{(I_{n}-\Theta)(I_{n}-W(x)^{T}\Theta)^{-1}}_{1}=1$and $\norm{I_{n}-C^{T}}_{1}=2 $, we have
\begin{equation*}
\begin{split}
\left\lVert\partial F/\partial x\right\rVert_{1}\leq &2\norm{\Theta(I_{n}-\Theta)^{-1}}_{1} \norm{\diag(F(x))}_{1}\\
=&\!\dfrac{2\thetamax}{1-\thetamax} \max_{i}F_{i}(x).
\end{split}
\end{equation*}
Since $\max_{i}F_{i}(x)\leq(1+\zeta)/n$ and $\thetamax<\dfrac{n}{n+2(1+\zeta)}$, we have 
\begin{equation*}
\left\lVert\dfrac{\partial F}{\partial x}\right\rVert_{1}\leq\dfrac{2\thetamax}{n(1-\thetamax)}(1+\zeta)<1.
\end{equation*}
Denote $\kappa=\dfrac{2\thetamax(1+\zeta)}{n(1-\thetamax)}$. Now, following the mean value inequality (Theorem 3.2.3, \cite{JMO-WCR:70}), we have that for any $y,z\in \interior\Delta_{n}$, there holds 
\begin{equation*}
\begin{split}
\norm{F(y)-F(z)}_{1}&\leq \sup_{0\leq t\leq1}\left\lVert \dfrac{\partial F}{\partial x}\bigg|_{x=z+t(y-z)}\right\rVert_{1} \norm{y-z}_{1}\\
&\leq \kappa\norm{y-z}_{1}<\norm{y-z}_{1},
\end{split}
\end{equation*}
i.e., $F(x)$ is contractive on $\interior\Delta_{n}$. Moreover, for any $y\in \Delta_{n}\setminus \interior\Delta_{n}$ and $z\in \Delta_{n}$, since $\Delta_{n}$ is compact, there exists a Cauchy sequence $\lbrace y_{k}\rbrace_{k=0}^{\infty}$, which satisfies $y_{k}\in \interior\Delta_{n}$ and $\lim_{k\to\infty}y_{k}=y$. Therefore,
\begin{equation*}
\begin{split}
&\norm{F(y)-F(z)}_{1}\\=&\norm{F(\lim_{k\to\infty}y_{k})-F(z)}_{1}\\
\leq &\lim_{k\to\infty}\sup_{0\leq t\leq1}\left\lVert \dfrac{\partial F}{\partial x}\bigg|_{x=z+t(y_{k}-z)}\right\rVert_{1} \norm{y_{k}-z}_{1}\\
\leq &\kappa\lim_{k\to\infty}\norm{y_{k}-z}_{1}=\kappa \norm{y-z}_{1}<\norm{y-z}_{1}.
\end{split}
\end{equation*}
Similarly, for any  $y,z\in \Delta_{n}\setminus \interior\Delta_{n}$, there holds $\norm{F(y)-F(z)}_{1}\leq\kappa\norm{y-z}_{1}<\norm{y-z}_{1}$. That is, for any $y,z\in\Delta_{n}$, there holds $\norm{F(y)-F(z)}_{1}<\norm{y-z}_{1}$. Thus, $F(x)$ is contractive on $\Delta_{n}$. Then the uniqueness of $x^{*}$ follows from the Banach fixed point theorem. 

\section{Proof of Corollary \ref{C1}}
Regarding (i), for any $i\in \Vfully$ and $j\in \Vpartly$, equation \eqref{e11} implies that
\begin{equation*}
\begin{split}
x_{i}^{*}-x_{j}^{*}=&\dfrac{\theta_{j}}{n}\!-\!\theta_{j}(x_{j}^{*})^{2}\!+\!C_{ji}\dfrac{\theta_{j}}{1-\theta_{j}}x_{j}^{*}(1-x_{j}^{*})\\
&+\theta_{j}\!\sum_{\mathclap{k\in \Vpartly\setminus\{j\}}}\!C_{kj}\dfrac{\theta_{k}}{1\!-\!\theta_{k}}x_{k}^{*}(1-x_{k}^{*}).\\
\end{split}
\end{equation*}
Note that $\sum_{k\in \Vpartly\setminus\{j\}}\!C_{kj}\dfrac{\theta_{k}}{1\!-\!\theta_{k}}x_{k}^{*}(1-x_{k}^{*})=\dfrac{1-\theta_{j}x_{j}}{1-\theta_{j}}x_{j}^{*}-\dfrac{1}{n}$. Then,
\begin{equation*}
\begin{split}
x_{i}^{*}-x_{j}^{*}\!=\!\theta_{j}(\dfrac{C_{ji}(1\!-\!x_{j}^{*})}{1-\theta_{j}}x_{j}^{*}\!+\!x_{j}^{*}(\dfrac{1\!-\!\theta_{j}x_{j}^{*}}{1-\theta_{j}}-x_{j}^{*}))\!>\!0,
\end{split}
\end{equation*}
Therefore, $x_{i}^{*}>x_{j}^{*}$.

Regarding (ii), for any $i,j\in \Vpartly$, by equation \eqref{e11},
\begin{equation*}
\begin{split}
&x_{i}^{*}-x_{j}^{*}=\theta_{i}(x_{i}^{*})^{2}\!-\!\theta_{j}(x_{j}^{*})^{2}\!+\!\dfrac{\theta_{j}\!-\!\theta_{i}}{n}\!\\
&+\!C_{ji}\!\dfrac{\theta_{j}\!(1-\theta_{i})}{1\!-\!\theta_{j}}x_{j}^{*}\!(1-x_{j}^{*})\!-\!C_{ij}\dfrac{\theta_{i}(1\!-\!\theta_{j})}{1\!-\!\theta_{i}}x_{i}^{*}(1\!-\!x_{i}^{*})\\
&+\!\sum_{\mathclap{k\in \Vpartly\setminus\{i,j\}}}\!C_{ki}\dfrac{\theta_{k}(\theta_{j}\!-\!\theta_{i})}{1\!-\!\theta_{k}}x_{k}^{*}(1\!-\!x_{k}^{*}).
\end{split}
\end{equation*}
Note that $\theta_{i}(x_{i}^{*})^{2}-\theta_{j}(x_{j}^{*})^{2}=(\theta_{i}-\theta_{j})(x_{i}^{*})^{2}+\theta_{j}(x_{i}^{*}-x_{j}^{*})(x_{i}^{*}+x_{j}^{*})$, $\sum_{k\in \Vpartly\setminus\{i,j\}}C_{ki}\dfrac{\theta_{k}}{1-\theta_{k}}x_{k}^{*}(1-x_{k}^{*})=\dfrac{1-\theta_{i}x_{i}^{*}}{1-\theta_{i}}x_{i}^{*}-\dfrac{1}{n}-\theta_{j}C_{ij}x_{j}^{*}\dfrac{1-x_{j}^{*}}{1-\theta_{j}}$, and $\theta_{j}C_{ji}x_{j}^{*}(1-x_{j}^{*})\dfrac{1-\theta_{i}}{1-\theta_{j}}-\theta_{i}C_{ij}x_{i}^{*}(1-x_{i}^{*})\dfrac{1-\theta_{j}}{1-\theta_{i}}=C_{ij}(\theta_{j}(x_{j}^{*}-x_{i}^{*})(1-x_{i}^{*}-x_{j}^{*})\dfrac{1-\theta_{i}}{1-\theta_{j}}+(1-\theta_{i}\theta_{j})x_{i}^{*}(1-x_{i}^{*})\dfrac{\theta_{j}-\theta_{i}}{(1-\theta_{i})(1-\theta_{j})})$. Then, we obtain
\begin{equation*}
\begin{split}
&(x_{i}^{*}-x_{j}^{*})(1-\theta_{j}(x_{i}^{*}+x_{j}^{*})+\theta_{j}C_{ij}(1-x_{i}^{*}-x_{j}^{*}))\\
=&(\theta_{j}-\theta_{i})(x_{i}^{*}(\dfrac{1-\theta_{i}x_{i}^{*}}{1-\theta_{i}}-x_{i}^{*})+C_{ij}\dfrac{x_{i}^{*}(1-x_{i}^{*})}{1-\theta_{i}}),
\end{split}
\end{equation*}
where $\theta_{i}>\theta_{j}$ and $1-\theta_{i}x_{i}^{*}>1-\theta_{i}$ indicate that the right hand side is negative. Moreover, since $x_{i}^{*}+x_{j}^{*}<1$, then $1-\theta_{j}(x_{i}^{*}+x_{j}^{*})+\theta_{j}C_{ij}(1-x_{i}^{*}-x_{j}^{*})>0$, which implies that $x_{i}^{*}<x_{j}^{*}$. 

Regarding (iii), for any $i,j$,
\begin{small}
\begin{equation*}
\begin{split}
&x_{i}^{*}-x_{j}^{*}-(\theta_{i}(x_{i}^{*})^{2}-\theta_{j}(x_{j}^{*})^{2})\\
=&\dfrac{\theta_{j}\!-\!\theta_{i}}{n}\!+\!C_{ji}\dfrac{\theta_{j}(\!1\!-\!\theta_{i}\!)}{1\!-\!\theta_{j}}x_{j}^{*}(1\!-\!x_{j}^{*})\!-\!C_{ij}\dfrac{\theta_{i}(\!1\!-\!\theta_{j}\!)}{1\!-\!\theta_{i}}x_{i}^{*}(1\!-\!x_{i}^{*})\\
+&\!(1\!-\!\theta_{i})\!\sum_{\mathclap{k\in \Vpartly\setminus\{i,j\}}}\!C_{ki}\dfrac{\theta_{k}x_{k}^{*}(1\!-\!x_{k}^{*})}{1-\theta_{k}}\!-\!(1\!-\!\theta_{j})\!\sum_{\mathclap{k\in \Vpartly\setminus\{i,j\}}}\!C_{kj}\dfrac{\theta_{k}x_{k}^{*}(1\!-\!x_{k}^{*})}{1-\theta_{k}}
\end{split}
\end{equation*}
\end{small}
Moreover, we have 
\begin{small}
\begin{equation*}
\begin{split}
&(1\!-\!\theta_{i})\!\sum_{\mathclap{k\in \Vpartly\setminus\{i,j\}}}C_{ki}\dfrac{\theta_{k}x_{k}^{*}(1\!-\!x_{k}^{*})}{1-\theta_{k}}\!-\!(1\!-\!\theta_{j})\sum_{\mathclap{k\in \Vpartly\setminus\{i,j\}}}C_{kj}\dfrac{\theta_{k}x_{k}^{*}(1\!-\!x_{k}^{*})}{1-\theta_{k}}\\
&=(\theta_{j}-\theta_{i})\sum_{\mathclap{k\in \Vpartly\setminus\{i,j\}}}C_{ki}\dfrac{\theta_{k}x_{k}^{*}(1-x_{k}^{*})}{1-\theta_{k}}\\
&+(1-\theta_{j})\sum_{\mathclap{k\in \Vpartly\setminus\{i,j\}}}\!(C_{ki}-C_{kj})\dfrac{\theta_{k}x_{k}^{*}(1-x_{k}^{*})}{1-\theta_{k}}\!,
\end{split}
\end{equation*}
\end{small}
and
\begin{equation*}
\begin{split}
&\!(1\!-\!\theta_{j})\!\sum_{\mathclap{k\in \Vpartly\setminus\{i,j\}}}\!(C_{ki}\!-\!C_{kj})\!\dfrac{\theta_{k}x_{k}^{*}(1\!-\!x_{k}^{*})}{1-\theta_{k}}\!=\\
&\!C_{ij}\theta_{j}(x_{i}^{*}\!-\!x_{j}^{*})(1\!-\!x_{j}^{*}\!-\!x_{i}^{*})\!+\!(x_{i}^{*}\!-\!x_{j}^{*})(1\!-\!\theta_{j}(x_{j}^{*}\!+\!x_{i}^{*})\!)\\
&+\dfrac{\theta_{i}\!-\!\theta_{j}}{1\!-\!\theta_{i}}x_{i}^{*}\!+\!\dfrac{\theta_{j}\!-\!\theta_{i}}{1\!-\!\theta_{i}}(x_{i}^{*})^{2}\!+\!C_{ij}\dfrac{\theta_{i}-\theta_{j}}{1-\theta_{i}}x_{i}^{*}(1\!-\!x_{i}^{*}).
\end{split}
\end{equation*}
Then it follows that 
\begin{equation*}
\begin{split}
&(x_{i}^{*}-x_{j}^{*})C_{ij}\theta_{j}(1-x_{j}^{*}-x_{i}^{*})(\dfrac{1-\theta_{i}}{1-\theta_{j}}-1)\\
=&(\theta_{j}\!-\!\theta_{i})(\dfrac{1}{n}\!-\!\dfrac{1\!-\!\theta_{i}x_{i}^{*}}{1-\theta_{i}}x_{i}^{*}\!+\!C_{ij}\dfrac{\theta_{j}x_{i}^{*}(1\!-\!x_{i}^{*})}{1-\theta_{j}}\!\\
&+\sum_{\mathclap{k\in \Vpartly\setminus\{i,j\}}}\!C_{ki}\dfrac{\theta_{k}x_{k}^{*}(1\!-\!x_{k}^{*})}{1-\theta_{k}}).
\end{split}
\end{equation*}
Since $\dfrac{1}{n}\!-\!\dfrac{1\!-\!\theta_{i}x_{i}^{*}}{1-\theta_{i}}x_{i}^{*}>0$, we obtain that $x_{i}^{*}<x_{j}^{*}$ holds if and only if $\theta_{i}>\theta_{j}$.

\section{Proof of Theorem \ref{T2}}
According to Lemma \ref{L4}, we just need to show the statements hold for system (\ref{e4}). Without loss of generality, let node $1$ be the center node. Then, $\Theta$ and $C$ can be written as
\begin{equation*}
\Theta=\begin{bmatrix}
 0_{r\times r} & 0 \\
  0 & \ \ \ \Theta_{p}
  \end{bmatrix}, \ \ {\rm and} \ \ C=\begin{bmatrix}
 C_{f} & C_{fp} \\
  C_{pf} & \ \ \ 0
  \end{bmatrix},
\end{equation*}
where $\Theta_{p}=\diag(\theta_{r+1},\theta_{r+2},...,\theta_{n})$, and $C_{pf}=\mathbf{1}_{n-r}\mathbf{e}_{1}^{T}$ with $\mathbf{e}_{1}$ being a $r$-dimensional vector whose first element is $1$ and others are $0$. Let $V(x)=(I_{n}-\Theta W(x))^{-1}(I_{n}-\Theta)$, we have
\begin{equation*}
\begin{split}
&V(x)\\
=&(I_{n}-\Theta \diag(x)-\Theta C+\Theta \diag(x) C)^{-1}(I_{n}-\Theta)\\
\!=\!&\begin{bmatrix}
 I_{r} & 0 \\
 -\Theta_{p}(I_{n-r}\!-\!\diag(x_{p})\!)C_{pf} & I_{n-r}\!-\!\Theta_{p}\!\diag(\!x_{p}\!)\!
  \end{bmatrix}^{-1}\\
  &\times\begin{bmatrix}
 I_{r} & 0 \\
  0 & \ \ I_{n-r}\!-\!\Theta_{p}
  \end{bmatrix}=\begin{bmatrix}
 I_{r} & 0 \\
 V_{pf}(x) & V_{p}(x)
  \end{bmatrix},
\end{split}
\end{equation*}
where $x_{p}=(x_{r+1},x_{r+2},...,x_{n})^{T}$, $V_{pf}(x)=(I_{n-r}-\Theta_{p}\diag(x_{p}))^{-1}\Theta_{p}(I_{n-r}-\diag(x_{p}))C_{pf}$, and $V_{p}(x)=(I_{n-r}-\Theta_{p}\diag(x_{p}))^{-1}(I_{n-r}-\Theta_{p})$. Therefore,
\begin{equation*}
\begin{split}
F(x)\!=\!\dfrac{1}{n}\begin{bmatrix}
\mathbf{1}_{r}\!+\!\mathbf{e}_{1}\mathbf{1}_{n-r}^{T}V^{'}\Theta_{p}(I_{n-r}-\diag(x_{p}))\mathbf{1}_{n-r}\\
 V^{'}(I_{n-r}-\Theta_{p})\mathbf{1}_{n-r}
 \end{bmatrix}, 
\end{split}
\end{equation*}
where $V^{'}=(I_{n-r}-\Theta_{p}\diag(x_{p}))^{-1}$. Note that $\mathbf{1}_{n-r}^{T}V^{'}\Theta_{p}(I_{n-r}-\diag(x_{p}))\mathbf{1}_{n-r}=\sum_{j=r+1}^{n}\dfrac{\theta_{j}(1-x_{j})}{1-\theta_{j}x_{j}}$, thus we obtain that 1) $F_{1}(x)=\dfrac{1}{n}+\sum_{j=r+1}^{n}\dfrac{\theta_{j}(1-x_{j})}{n(1-\theta_{j}x_{j}})$; 2) $F_{i}(x)=\dfrac{1}{n}$ for $i\in \Vfully\setminus\{1\}$; 3) $F_{i}(x)=\dfrac{1-\theta_{i}}{n(1-\theta_{i}x_{i})}$ for $i\in\Vpartly$.

By Theorem \ref{T1}, we have $x^{*}_{i}\in\interior\Delta_{n}$. Regarding (ii), since $F_{i}(x)=1/n$ for $i\in \Vfully\setminus\{1\}$, then $x^{*}_{i}=1/n$ for any $i\in \Vfully\setminus\{l\}$. Regarding (iii), for $i\in \Vpartly$, we have that $F_{i}(x)=F_{i}(x_{i})=\dfrac{1-\theta_{i}}{n(1-\theta_{i}x_{i})}$. Since $F_{j}(x)\geq 1/n$ for any $j\in \Vfully$, $F_{i}(x)\in[0,1-r/n]$. Next, we show that $F_{i}(x_{i})$ is contractive on $[0,1-r/n]$. For any $x_{i}^{'},x_{i}^{''}\in[0,1-r/n]$,
\begin{equation*}
\begin{split}
&\mid F_{i}(x_{i}^{'})-F_{i}(x_{i}^{''})\mid=\dfrac{\theta_{i}(1-\theta_{i})\mid x_{i}^{'}-x_{i}^{''}\mid}{n(1-\theta_{i}x_{i}^{'})(1-\theta_{i}x_{i}^{''})}\\
\leq&\dfrac{\theta_{i}(1-\theta_{i})}{n}\dfrac{\mid x_{i}^{'}-x_{i}^{''}\mid}{(1-\theta_{i}(1-r/n))^{2}}<\mid x_{i}^{'}-x_{i}^{''}\mid,
\end{split}
\end{equation*}
in which $\dfrac{\theta_{i}(1-\theta_{i})}{n(1-\theta_{i}(1-r/n))^{2}}<1$ follows from the fact that $(n+\dfrac{r(r-2)}{n}-1)\theta_{i}^{2}-2(n-r-\dfrac{1}{2})\theta_{i}+n>0$ for any $0<\theta_{i}<1$. Therefore, $F_{i}(x_{i})$ is contractive on $[0,1-r/n]$ for any $i\in \Vpartly$. By the Banach fixed point theorem, $x_{i}(s)$ globally converges to unique equilibrium for any $x(0)\in \Delta_{n}$. In conclusion, for any $i\neq l$, $x_{i}(s)$ converges to unique $x^{*}_{i}$. Moreover, for any $i\in \Vpartly$, by the proof of Theorem \ref{T1}, we obtain $x^{*}_{i}=\dfrac{n-\sqrt{n^{2}-4n\theta_{i}(1-\theta_{i})}}{2n\theta_{i}}$. Suppose that $x^{*}_{i}$ is non-decreasing with respect to $\theta_{i}$. Then, taking the derivative of $x^{*}_{i}$ with respect to $\theta_{i}$, we obtain
\begin{equation*}
\dfrac{2n\theta_{i}-4n\theta_{i}^{2}}{\sqrt{n^{2}-4n\theta_{i}(1-\theta_{i})}}-n+\sqrt{n^{2}-4n\theta_{i}(1-\theta_{i})}\geq 0,
\end{equation*}
which indicates that $n-2\theta_{i}-\sqrt{n^{2}-4n\theta_{i}(1-\theta_{i})}\geq0$. Since $n-2\theta_{i}>0$, we have $(n-2\theta_{i})^{2}\geq n^{2}-4n\theta_{i}(1-\theta_{i})$, i.e., $1\geq n$, which is a contradiction. Thus, $x^{*}_{i}$ is decreasing with respect to $\theta_{i}$. Regarding (iv), since $x_{l}(s+1)=1/n+\sum_{j=r+1}^{n}\dfrac{\theta_{j}(1-x_{j}(s))}{n(1-\theta_{j}x_{j}(s))}$, then $x_{l}(s)$ globally converges to $\dfrac{1}{n}+\sum_{j=r+1}^{n}\dfrac{\theta_{j}(1-x_{j}^{*})}{n(1-\theta_{j}x_{j}^{*})}>\dfrac{1}{n}$.

\section{Proof of Theorem \ref{T3}}
Let $\beta_{i}=\theta_{i}(1-x_{i})$ and $\gamma_{i}=1-\theta_{i}x_{i}$. Without loss of generality, Let node $r+1$ be the center node of $\mathcal{G}(C)$, i.e., $l=r+1$. Similarly, $C$ can be written as
\begin{equation*}
C=\begin{bmatrix}
 0 & C_{fp} \\
  C_{pf} & \ \ \ C_{p}
  \end{bmatrix},
\end{equation*}
where $C_{pf}=\mathbf{e}_{1}(C_{l1}, C_{l2},...,C_{lr})$, $C_{p}=(\mathbf{1}_{n-r}-\mathbf{e}_{1})\mathbf{e}_{1}^{T}+\mathbf{e}_{1}(0, C_{ll+1},...,C_{ln})$ with $\mathbf{e}_{1}$ being a $(n-r)$-dimensional vector whose first element is $1$ and others are $0$. Then,
\begin{equation*}
\begin{split}
&V(x)\\
=&\begin{bmatrix}
 I_{r\times r} & 0 \\
 \hat{V}^{-1}\Theta_{p}(I_{n-r}-\diag(x_{p}))C_{pf} & \hat{V}^{-1}(I_{n-r}-\Theta_{p})
  \end{bmatrix},
  \end{split}
\end{equation*}
where $x_{p}=(x_{r+1},x_{r+2},...,x_{n})^{T}$, $\hat{V}=I_{n-r}-\Theta_{p}\diag(x_{p})-\Theta_{p}(I_{n-r}-\diag(x_{p}))C_{p}$. Therefore,
\begin{equation*}
\begin{split}
F(x)\!=\!\dfrac{1}{n}\!\begin{bmatrix}
\mathbf{1}_{r}\!+\!C_{pf}^{T}\Theta_{p}(I_{n-r}\!-\!\diag(x_{p}))(\hat{V}^{-1})^{T}\mathbf{1}_{n-r}\\
 (I_{n-r}-\Theta_{p})(\hat{V}^{-1})^{T}\mathbf{1}_{n-r}
 \end{bmatrix}. 
\end{split}
\end{equation*}
By column transformations, we obtain the first column of $\hat{V}^{-1}$ is $\dfrac{\vartheta}{\alpha}$, and the $i$-th column of $\hat{V}^{-1}$ is $\dfrac{\mathbf{e}_{i}}{\gamma_{i+r}}+\dfrac{\vartheta\beta_{r+1}C_{l \ r+i}}{\alpha\gamma_{r+i}}$ for $2\leq i\leq n-r$, where $\vartheta=(1,\dfrac{\beta_{r+2}}{\gamma_{r+2}},\dfrac{\beta_{r+3}}{\gamma_{r+3}},...,\dfrac{\beta_{n}}{\gamma_{n}})^{T}$, $\alpha=\gamma_{l}-\beta_{l}\sum_{j\in \Vpartly\setminus\{l\}}C_{lj}\dfrac{\beta_{j}}{\gamma_{j}}$. Thus, we have that $F_{i}(x)=\dfrac{\xi(1-\theta_{l})}{n\alpha}$ for $i=l$, $F_{i}(x)=\dfrac{1}{n}+\xi\dfrac{\beta_{l}C_{li}}{n\alpha}$ for any $i\in \Vfully$, and $F_{i}(x)==\dfrac{1-\theta_{i}}{n\gamma_{i}}+\xi\dfrac{\beta_{l}C_{li}(1-\theta_{i})}{n\alpha\gamma_{i}}$ for any $i\in\Vpartly\setminus\{l\}$, where $\xi=1+\sum_{j\in \Vpartly\setminus\{l\}}\dfrac{\beta_{j}}{\gamma_{j}}$.

Similarly, we only need to show that the equilibrium social power of system (\ref{e4}) satisfies all statements. Denote $\beta_{i}^{*}$, $\gamma_{i}^{*}$, $\alpha^{*}$ and $\xi^{*}$ as $\beta_{i}$, $\gamma_{i}$, $\alpha$ and $\xi$ corresponding to $x^{*}$, respectively. 
Regarding (i), a) by Theorem \ref{T1}, we have $x^{*}\in\interior\Delta_{n}$. 
b) For $i\in \Vfully$, we have that $x^{*}_{i}=\dfrac{1}{n}+\xi^{*}\dfrac{\beta_{l}^{*}C_{li}}{n\alpha^{*}}$. Then, if $C_{li}=0$, $x^{*}_{i}=\dfrac{1}{n}$. Otherwise, $x^{*}_{i}>\dfrac{1}{n}$ follows from that $\alpha^{*}$, $\xi^{*}$ and $\beta_{l}^{*}$ are all positive. 
c) For $i\in \Vpartly\setminus\{l\}$, if $C_{li}=0$, we have that $x^{*}_{i}=\dfrac{1-\theta_{i}}{n(1-\theta_{i}x^{*}_{i})}$. Then, by the proof of Theorem \ref{T2} we have that $x^{*}_{i}=\dfrac{n-\sqrt{n^{2}-4n\theta_{i}(1-\theta_{i})}}{2n\theta_{i}}$ and is decreasing with respect to $\theta_{i}$. 

Regarding (ii), since $C_{li}=0$ for any $i\in \Vpartly\setminus\{l\}$, we have that $\alpha(s)=\gamma_{l}(s)$ and $\alpha^{*}=\gamma^{*}_{l}$. a) For $i\in \Vpartly\setminus\{l\}$, since $C_{li}=0$, by Theorem \ref{T2} $x_{i}(s)$ globally converges to $x^{*}_{i}=\dfrac{1-\theta_{i}}{n(1-\theta_{i}x^{*}_{i})}$. 
b) Note that $x^{*}_{l}=\dfrac{\xi^{*}(1-\theta_{l})}{n\gamma^{*}_{l}}$, then $x^{*}_{l}=\dfrac{n-\sqrt{n^{2}-4n\theta_{l}(1-\theta_{l})\xi^{*}}}{2n\theta_{l}}$ since $x^{*}_{l}<1$. Moreover, for $i\in \Vpartly\setminus\{l\}$, since $C_{li}=0$, we have that $nx^{*}_{i}=\dfrac{1-\theta_{i}}{\gamma^{*}_{i}}$, which implies that $\dfrac{\beta_{i}^{*}}{\gamma^{*}_{i}}=1-\dfrac{1-\theta_{i}}{\gamma^{*}_{i}}=1-nx^{*}_{i}$. Therefore, $\xi^{*}=1+\sum_{j\in \Vpartly\setminus\{l\}}\beta_{j}^{*}/\gamma^{*}_{j}=1+\sum_{j\in \Vpartly\setminus\{l\}}(1-nx^{*}_{i})=n-r-n\sum_{j\in \Vpartly\setminus\{l\}}x^{*}_{j}$. Then, the uniqueness of $x^{*}_{l}$ follows from the uniqueness of $\xi^{*}$.
c) For $i\in \Vfully$, we have $x^{*}_{i}=\dfrac{1}{n}+\dfrac{\xi^{*}(\gamma^{*}_{l}-(1-\theta_{l}))}{n\gamma^{*}_{l}}C_{li}$. Since $x^{*}_{l}=\dfrac{\xi^{*}(1-\theta_{l})}{n\gamma^{*}_{l}}$, we have $x^{*}_{i}=\dfrac{1}{n}+(\dfrac{\xi^{*}}{n}-x^{*}_{l})C_{li}$. Finally, the uniqueness of $x^{*}_{i}$ follows from the fact that $\xi^{*}$ and $x^{*}_{l}$ are both unique.

\section{Proof of Corollary \ref{C3}}
First, we show that the equilibrium social power of system (\ref{e4}) satisfies all statements. Regarding (i), for any $i,j\in \Vfully$, since $\alpha^{*}$, $\xi^{*}$ and $\beta_{l}^{*}$ are all positive, we have that $x^{*}_{i}=\dfrac{1}{n}+\xi^{*}\dfrac{\beta_{l}^{*}C_{li}}{n\alpha^{*}}>\dfrac{1}{n}+\xi^{*}\dfrac{\beta_{l}^{*}C_{lj}}{n\alpha^{*}}=x^{*}_{j}$ if and only if $C_{li}>C_{lj}$. 

Regarding (ii), For any $i\in \Vfully$ and $j\in \Vpartly\setminus\{l\}$ with $C_{li}=C_{lj}$,
\begin{equation*}
\begin{split}
x^{*}_{i}-x^{*}_{j}&=\dfrac{1}{n}(1-\dfrac{1-\theta_{j}}{\gamma_{j}^{*}})+\dfrac{\xi^{*}\beta_{l}^{*}C_{lj}}{n\alpha^{*}}(1-\dfrac{1-\theta_{j}}{\gamma_{j}^{*}})\\
&=\dfrac{1}{n}(1+\dfrac{\xi^{*}\beta_{l}^{*}C_{lj}}{n\alpha^{*}})(1-\dfrac{1-\theta_{j}}{\gamma_{j}^{*}}).
  \end{split}
\end{equation*}
Since $1+\dfrac{\xi^{*}\beta_{l}^{*}C_{lj}}{n\alpha^{*}}>0$ and $1-\dfrac{1-\theta_{j}}{\gamma_{j}^{*}}>0$, we obtain that $x^{*}_{i}-x^{*}_{j}>0$.

Regarding (iii), for any $i,j\in \Vpartly\setminus\{l\}$ with $C_{li}=C_{lj}$, 
\begin{equation*}
\begin{split}
x^{*}_{i}-x^{*}_{j}=&\dfrac{1}{n}(1+\xi^{*}\dfrac{\beta^{*}_{l}C_{li}}{\alpha^{*}})(\dfrac{1-\theta_{i}}{\gamma^{*}_{i}}-\dfrac{1-\theta_{j}}{\gamma^{*}_{j}})\\
=&\dfrac{1-\theta_{j}}{n}(1+\xi^{*}\dfrac{\beta^{*}_{l}C_{lj}}{\alpha^{*}})(\dfrac{1}{\gamma^{*}_{i}}-\dfrac{1}{\gamma^{*}_{j}})\\
&-\dfrac{1}{n\gamma^{*}_{i}}(1+\xi^{*}\dfrac{\beta^{*}_{l}C_{li}}{\alpha^{*}})(\theta_{i}-\theta_{j}),
\end{split}
\end{equation*}
where the last equation follows from that $C_{li}=C_{lj}$ and $\dfrac{\theta_{i}}{\gamma^{*}_{i}}-\dfrac{\theta_{j}}{\gamma^{*}_{j}}=\dfrac{1}{\gamma^{*}_{i}}(\theta_{i}-\theta_{j})+\theta_{j}(\dfrac{1}{\gamma^{*}_{i}}-\dfrac{1}{\gamma^{*}_{j}})$. Note that $\dfrac{1-\theta_{j}}{n}(1+\xi^{*}\dfrac{\beta^{*}_{l}C_{lj}}{\alpha^{*}})=x^{*}_{j}\gamma^{*}_{j}$ and $\dfrac{1}{n\gamma^{*}_{i}}(1+\xi^{*}\dfrac{\beta^{*}_{l}C_{li}}{\alpha^{*}})=\dfrac{x^{*}_{i}}{1-\theta_{i}}$, thus, $x^{*}_{i}-x^{*}_{j}=x^{*}_{j}\gamma^{*}_{j}(\dfrac{1}{\gamma^{*}_{i}}-\dfrac{1}{\gamma^{*}_{j}})-\dfrac{x^{*}_{i}}{1-\theta_{i}}(\theta_{i}-\theta_{j})$. That is, 
\begin{equation*}
\begin{split}
&-x^{*}_{i}\dfrac{\gamma^{*}_{i}}{1-\theta_{i}}(\theta_{i}-\theta_{j})=\gamma^{*}_{i}x^{*}_{i}-\gamma^{*}_{j}x^{*}_{j}\\
=&(x^{*}_{i}-x^{*}_{j})(1-\theta_{j}(x^{*}_{i}+x^{*}_{j}))-(x^{*}_{i})^2(\theta_{i}-\theta_{j}).
  \end{split}
\end{equation*}
Therefore,
\begin{equation*}
\begin{split}
(x^{*}_{i}\!-\!x^{*}_{j})(1-\theta_{j}(x^{*}_{i}+x^{*}_{j}))\!=\!x^{*}_{i}(x^{*}_{i}-\dfrac{\gamma^{*}_{i}}{1-\theta_{i}})(\theta_{i}-\theta_{j}).
  \end{split}
\end{equation*}
Since $x^{*}\in \interior\Delta_{n}$ and $\theta_{i},\theta_{j}<1$, we have that $1-\theta_{j}(x^{*}_{i}+x^{*}_{j})>0$ and $\gamma^{*}_{i}/(1-\theta_{i})>1>x^{*}_{i}$, which means that $x^{*}_{i}>x^{*}_{j}$ if and only if $\theta_{i}<\theta_{j}$.

Regarding (iv), for any $i,j\in \Vpartly\setminus\{l\}$ with $\theta_{i}=\theta_{j}$, we have that 
\begin{equation*}
\begin{split}
x^{*}_{i}-x^{*}_{j}&=\dfrac{1-\theta_{i}}{n}((\dfrac{1}{\gamma^{*}_{i}}-\dfrac{1}{\gamma^{*}_{j}})+\dfrac{\xi^{*}\beta^{*}_{l}}{\alpha^{*}}(\dfrac{C_{li}}{\gamma^{*}_{i}}-\dfrac{C_{lj}}{\gamma^{*}_{j}}))\\
&=x^{*}_{j}\gamma^{*}_{j}\dfrac{1}{\gamma^{*}_{i}}-x^{*}_{j}+\dfrac{(1-\theta_{i})\xi^{*}\beta^{*}_{l}}{n\alpha^{*}\gamma^{*}_{i}}(C_{li}-C_{lj}).
  \end{split}
\end{equation*}
Thus, it follows that
\begin{equation*}
\begin{split}
(x^{*}_{i}-x^{*}_{j})(1-\theta_{j}(x^{*}_{i}+x^{*}_{j}))\!=\!\dfrac{(1-\theta_{i})\xi^{*}\beta^{*}_{l}}{n\alpha^{*}}(C_{li}-C_{lj})
  \end{split}
\end{equation*}
with $\dfrac{(1-\theta_{i})\xi^{*}\beta^{*}_{l}}{n\alpha^{*}}>0$, which means that $x^{*}_{i}>x^{*}_{j}$ if and only if $C_{li}>C_{lj}$.

\section{Proof of Theorem \ref{T6}}
The proof of (i) follows from Theorem \ref{T2}. Regarding (ii), for $i\in \Vpartly\setminus\{l\}$, Theorem \ref{T2} implies that $x_{i}(s)$ converges to $x^{*}_{i}$ for any $x(0)\in \Delta_{n}$. Since $\xi(s)=1+\sum_{j\in \Vpartly\setminus\{l\}}(1-\dfrac{1-\theta_{j}}{\gamma_{j}(s)})$ only depends on $x_{j}(s)$, where $j\in \Vpartly\setminus\{l\}$, thus, $\xi(s)$ converges to $\xi^{*}$ for any $x(0)\in \Delta_{n}$. For $i\in \Vfully$, note that $x_{i}(s+1)=\dfrac{1}{n}+\dfrac{\xi(s)\beta_{l}(s)C_{li}}{n\gamma_{l}(s)}$, which depends on $\xi(s)$ and $x_{l}(s)$. Therefore, $x_{i}(s)$ converges if $x_{l}(s)$ converges for any $x(0)\in \Delta_{n}$. Since $F_{i}(x)\geq 1/n$ for $i\in{V}_{f}$ and $F_{i}(x)\geq(1-\theta_{i})/n$ for $i\in{V}_{p}$, we have $x(s)\in\{x\in \Delta_{n}\mid(1-\theta_{i})/n\leq x_{i}\leq a_{i}, i\in\Vpartly\}$ for any $s\geq1$, where $a_{i}=\dfrac{n-r}{n}-\sum_{j\in \Vpartly\setminus\{i\}}\dfrac{1-\theta_{j}}{n}$. Next, we show that $F_{l}(x)=\dfrac{1-\theta_{l}}{n\gamma_{l}}\xi$ is contractive on $\{x\in \Delta_{n}\mid(1-\theta_{i})/n\leq x_{i}\leq a_{i}, i\in\Vpartly\}$. Consider $x^{'},x^{''}\in\{x\in \Delta_{n}\mid(1-\theta_{i})/n\leq x_{i}\leq a_{i}, i\in\Vpartly\}$, we have that
\begin{equation*}
\begin{split}
&\mid F_{l}(x^{'})-F_{l}(x^{''})\mid=\dfrac{1\!-\!\theta_{l}}{n}\mid\dfrac{\xi^{'}}{\gamma^{'}_{l}}\!-\!\dfrac{\xi^{''}}{\gamma^{''}_{l}}\mid\\
\leq&\dfrac{1-\theta_{l}}{n\gamma^{'}_{l}}\mid\xi^{'}-\xi^{''}\mid+\xi^{''}\dfrac{1-\theta_{l}}{n}\mid\dfrac{1}{\gamma^{'}_{l}}-\dfrac{1}{\gamma^{''}_{l}}\mid.
\end{split}
\end{equation*}
On one hand,
\begin{equation*}
\begin{split}
\mid\xi^{'}-\xi^{''}\mid&=\mid\sum_{\mathclap{j\in \Vpartly\setminus\{l\}}}(\dfrac{1-\theta_{j}}{\gamma^{''}_{j}}-\dfrac{1-\theta_{j}}{\gamma^{'}_{j}})\mid\\
&\leq\sum_{\mathclap{j\in \Vpartly\setminus\{l\}}}(1-\theta_{j})\mid\dfrac{1}{\gamma^{'}_{j}}-\dfrac{1}{\gamma^{''}_{j}}\mid.
\end{split}
\end{equation*}
Because $a_{j}<1-r/n$ for $j\in \Vpartly\setminus\{l\}$, by Theorem \ref{T1}, we have that for any $j\in \Vpartly\setminus\{l\}$, 
\begin{equation*}
\begin{split}
\mid\dfrac{1}{\gamma^{'}_{j}}\!-\!\dfrac{1}{\gamma^{''}_{j}}\mid=\!\dfrac{\theta_{j}}{\gamma^{'}_{j}\gamma^{''}_{j}}\mid\!x^{'}_{j}\!-\!x^{''}_{j}\!\mid\leq\dfrac{\theta_{j}\mid\!x^{'}_{j}\!-\!x^{''}_{j}\!\mid}{(1\!-\!\theta_{j}(1\!-\!r/n))^{2}}.
\end{split}
\end{equation*}
Therefore, there holds
\begin{equation*}
\begin{split}
&\mid\xi^{'}-\xi^{''}\mid\\
\leq&\sum_{j\in \Vpartly\setminus\{l\}}\dfrac{\theta_{j}(1-\theta_{j})}{(1-\theta_{j}(1-r/n))^{2}}\mid x^{'}_{j}-x^{''}_{j}\mid\\
\leq&\max_{j\in \Vpartly\setminus\{l\}}\dfrac{\theta_{j}(1-\theta_{j})}{(1-\theta_{j}(1-r/n))^{2}}\sum_{\mathclap{j\in \Vpartly\setminus\{l\}}}\mid x^{'}_{j}-x^{''}_{j}\mid.
\end{split}
\end{equation*}
On the other hand, for the center node, 
\begin{equation*}
\begin{split}
\mid\dfrac{1}{\gamma^{'}_{l}}\!-\!\dfrac{1}{\gamma^{''}_{l}}\mid=\!\dfrac{\theta_{l}}{\gamma^{'}_{l}\gamma^{''}_{l}}\mid x^{'}_{l}\!-\!x^{''}_{l}\mid\leq\dfrac{\theta_{l}}{(1\!-\!\theta_{l}a_{l})^{2}}\mid x^{'}_{l}\!-\!x^{''}_{l}\mid.
\end{split}
\end{equation*}
Thus,
\begin{equation*}
\begin{split}
&\mid F_{l}(x^{'})-F_{l}(x^{''})\mid\leq\xi^{''}\dfrac{\theta_{l}(1-\theta_{l})}{n(1-\theta_{l}a_{l})^{2}}\mid x^{'}_{l}-x^{''}_{l}\mid+\\
&\dfrac{1\!-\!\theta_{l}}{n\gamma^{'}_{l}}\max_{j\in \Vpartly\setminus\{l\}}\dfrac{\theta_{j}(1-\theta_{j})}{(1\!-\!\theta_{j}(1-r/n))^{2}}\sum_{j\in \Vpartly\setminus\{l\}}\mid x^{'}_{j}\!-\!x^{''}_{j}\mid.
\end{split}
\end{equation*}
Denote $\lambda$ by
\begin{equation*}
\max\{\dfrac{1\!-\!\theta_{l}}{n\gamma^{'}_{l}}\max_{j\in \Vpartly\setminus\{l\}}\dfrac{\theta_{j}(1-\theta_{j})}{(1\!-\!\theta_{j}(1\!-\!r/n))^{2}},\dfrac{\xi^{''}\theta_{l}(1-\theta_{l})}{n(1\!-\!\theta_{l}a_{l})^{2}}\}.
\end{equation*}
Then, we have 
\begin{equation*}
\begin{split}
&\mid F_{l}(x^{'})-F_{l}(x^{''})\mid\\
\leq&\lambda\sum_{j\in \Vpartly}\mid x^{'}_{j}-x^{''}_{j}\mid\leq\lambda\parallel x^{'}-x^{''}\parallel_{1}.
\end{split}
\end{equation*}
By the proof of Theorem \ref{T1}, we have that $\dfrac{\theta_{j}(1-\theta_{j})}{(1\!-\!\theta_{j}(1\!-\!r/n))^{2}}<n$ for any $j\in \Vpartly\setminus\{l\}$, i.e., $\max_{j\in \Vpartly\setminus\{l\}}\dfrac{\theta_{j}(1-\theta_{j})}{(1\!-\!\theta_{j}(1\!-\!r/n))^{2}}<n$. Moreover, since $\dfrac{1\!-\!\theta_{l}}{\gamma^{'}_{l}}<1$, we have that $\dfrac{1\!-\!\theta_{l}}{n\gamma^{'}_{l}}\max_{j\in \Vpartly\setminus\{l\}}\dfrac{\theta_{j}(1-\theta_{j})}{(1\!-\!\theta_{j}(1\!-\!r/n))^{2}}<1$. Next, we prove that $\dfrac{\xi^{''}\theta_{l}(1-\theta_{l})}{n(1\!-\!\theta_{l}a_{l})^{2}}<1$ for any if $\sum_{j\in \Vpartly\setminus\{l\}}\theta_{j}\leq \dfrac{4n}{5}-1$. Note that $\dfrac{\xi^{''}\theta_{l}(1-\theta_{l})}{n(1\!-\!\theta_{l}a_{l})^{2}}<1$ means $(na_{l}^{2}+\xi^{''})\theta^{2}_{l}-(2na_{l}+\xi^{''})\theta_{l}+n>0$. That is, $(2na_{l}+\xi^{''})^{2}-4n(na_{l}^{2}+\xi^{''})<0$, which is equivalent to that $\xi^{''}<4n(1-a_{l})=4(n-1)-4\sum_{j\in \Vpartly\setminus\{l\}}\theta_{j}$. Since $\xi^{''}< 1+\sum_{j\in \Vpartly\setminus\{l\}}\theta_{j}$, and $\sum_{j\in \Vpartly\setminus\{l\}}\theta_{j}\leq\dfrac{4n}{5}-1$, we have that $\xi^{''}<4(n-1)-4\sum_{j\in \Vpartly\setminus\{l\}}\theta_{j}$. In conclusion, for any $x^{'},x^{''}\in\{x\in \Delta_{n}\mid(1-\theta_{i})/n\leq x_{i}\leq a_{i}, i\in\Vpartly\}$, we have that $\mid F_{l}(x^{'})-F_{l}(x^{''})\mid \leq\lambda\parallel x^{'}-x^{''}\parallel_{1}<\parallel x^{'}-x^{''}\parallel_{1}$, which means $F_{l}(x)$ is contractive on $\{x\in \Delta_{n}\mid(1-\theta_{i})/n\leq x_{i}\leq a_{i}, i\in\Vpartly\}$. By the Banach fixed point theorem, we have $x_{l}(s)$ converges for any $x(0)\in \Delta_{n}$, which implies that $x(s)$ globally converges to $x^{*}$ for any $x(0)\in \Delta_{n}$ exponentially fast.

\section{Proof of Theorem \ref{T8}}
Since $\hat{G}(x)$ is an analytic function of $x\in\mathcal{A}$, it is differentiable on $\interior\mathcal{A}$ and continuous on $\mathcal{A}$. Let $B=\Theta(I_{n}-C)V$ and $B_{i}^{T}$ be the $i$-th row of $B$. Then, 
\begin{equation*}
I_{n}\otimes \Theta W(\omega)x=\begin{bmatrix}
\Theta C V_{1} \\
\Theta C V_{2} \\
\vdots \\
\Theta C V_{n}
  \end{bmatrix}+\begin{bmatrix}
\diag(\omega)\Theta(I_{n}-C)V_{1} \\
\diag(\omega)\Theta(I_{n}-C)V_{2} \\
\vdots \\
\diag(\omega)\Theta(I_{n}-C)V_{n}
  \end{bmatrix}.
\end{equation*}
Furthermore, since $\partial \omega/\partial V_{i}=\mathbf{e}_{i}\mathbf{1}_{n}^{T}/n$, we obtain
\begin{equation*}
\dfrac{\partial (\diag(\omega)\Theta(I_{n}\!-\!C)V_{i})}{\partial V_{i}}\!=\!\diag(\omega)\Theta(I_{n}\!-\!C)+\dfrac{B_{ii}}{n}\mathbf{e}_{i}\mathbf{1}_{n}^{T},
\end{equation*}
and
\begin{equation*}
\dfrac{\partial (\diag(\omega)\Theta(I_{n}\!-\!C)V_{j})}{\partial V_{i}}=\dfrac{B_{ij}}{n}\mathbf{e}_{i}\mathbf{1}_{n}^{T}
\end{equation*}
for any $j\neq i$, where $B_{ij}$ is the $ij$-th entry of $B=\Theta(I_{n}-C)V$. Hence, 
\begin{equation*}
\dfrac{\partial (I_{n}\otimes\Theta W(\omega)x)}{\partial V_{i}}=\begin{bmatrix}
0_{n\times n} \\
0_{n\times n} \\
\vdots \\
\Theta C\\
\vdots \\
0_{n\times n}
  \end{bmatrix}\!+\!\begin{bmatrix}\!
B_{i1}\mathbf{e}_{i}\mathbf{1}_{n}^{T}/n \\
B_{i2}\mathbf{e}_{i}\mathbf{1}_{n}^{T}/n \\
\vdots \\
B_{ii}^{'}+\!B_{ii}\mathbf{e}_{i}\mathbf{1}_{n}^{T}/n\\
\vdots \\
B_{in}\mathbf{e}_{i}\mathbf{1}_{n}^{T}/n
  \end{bmatrix},
\end{equation*}
where $B_{ii}^{'}=\diag(\omega)\Theta(I_{n}-C)$. Consequently, $\partial \hat{G}/\partial x=I_{n}\otimes \Theta W(\omega)+H/n$, where $H=[B_{1}\otimes(\mathbf{e}_{1}\mathbf{1}_{n}^{T}) \ B_{2}\otimes(\mathbf{e}_{2}\mathbf{1}_{n}^{T}) \ ... \ B_{n}\otimes(\mathbf{e}_{n}\mathbf{1}_{n}^{T})]$. Note that 
\begin{equation*}
\begin{split}
\left\lVert\dfrac{\partial \hat{G}}{\partial x}\right\rVert_{\infty}\!\leq\norm{\Theta W(x)}_{\infty}\!+\!\dfrac{\norm{H}_{\infty}}{n}\!=\!\thetamax+\!\max\limits_{i,j}\mid B_{ij}\mid.
\end{split}
\end{equation*}
Since $0\leq V_{ij}\leq 1$ and $0 \leq\sum_{k=1}^{n}C_{ik}V_{kj}\leq 1$, we have $\mid V_{ij}-\sum_{k=1}^{n}C_{ik}V_{kj} \mid\leq V_{ij}\leq 1$ if $V_{ij}\geq\sum_{k=1}^{n}C_{ik}V_{kj}$, and $\mid V_{ij}-\sum_{k=1}^{n}C_{ik}V_{kj} \mid\leq\sum_{k=1}^{n}C_{ik}V_{kj}\leq 1$ if $V_{ij}\leq\sum_{k=1}^{n}C_{ik}V_{kj}$. Thus, it follows that $\mid B_{ij} \mid=\theta_{i} \mid V_{ij}-\sum_{k=1}^{n}C_{ik}V_{kj} \mid \leq\thetamax$. Therefore, we obtain that $\partial \hat{G}/\partial x\leq 2\thetamax<1$. Similar with the proof of Theorem \ref{T1}, we obtain that $\hat{G}(x)$ is contracitve on $\mathcal{A}$. Then, exponential convergence of system (\ref{e14}) follows from the Banach fixed point theorem.

\end{document}